\documentclass{aa}
\usepackage{psfig}
\usepackage{rotating}
\newcommand{\beq}{\begin{equation}}
\newcommand{\eeq}{\end{equation}}
\newcommand{\gapprox}{$\stackrel {>}{_{\sim}}$}
\newcommand{\lapprox}{$\stackrel {<}{_{\sim}}$}

\def\farcm{\hbox{$.\mkern-4mu^\prime$}}

\def\arcmin{\hbox{$^\prime$}}
\def\arcsec{\hbox{$^{\prime\prime}$}}
\def\solar{\mbox{$_{\normalsize\odot}$}}

\begin{document}
\title{Mass Segregation in Young Magellanic Clouds Star
Clusters: Four Clusters observed with HST.}

\author{D. Gouliermis\inst{1,2}
\and S. C. Keller\inst{3}
\and M. Kontizas\inst{4}
\and E. Kontizas\inst{5}
\and I. Bellas-Velidis\inst{5}
}

\offprints{D. Gouliermis\\
\email{dgoulier@mpia.de}}

\institute{Sternwarte der Universit\"{a}t Bonn, Auf dem H\"{u}gel
        71, D-53121 Bonn, Germany
        \and Max-Planck-Institut f\"{u}r Astronomie, K\"{o}nigstuhl 17, 
         D-69117 Heidelberg, Germany
        \and Research School of Astronomy and Astrophysics, Mount Stromlo
Observatory, Weston, A.C.T. 2611, Australia
        \and Department of Astrophysics Astronomy \& Mechanics, Faculty
of Physics, University of Athens, GR-15783 Athens, Greece
        \and Institute for Astronomy and Astrophysics, National
Observatory of Athens, P.O. Box 20048, GR-11810 Athens, Greece
}

\date{Received .../ Accepted ...}

\abstract{ 
We present the results of our investigation on the phenomenon of mass
segregation in young star clusters in the Magellanic Clouds.  HST/WFPC2
observations on NGC 1818, NGC 2004 \& NGC 2100 in the Large Magellanic
Cloud and NGC 330 in the Small Magellanic Cloud have been used for the
application of diagnostic tools for mass segregation: i)  the radial
density profiles of the clusters for various mass groups and ii) their
mass functions (MFs) at various radii around their centres. All four
clusters are found to be mass segregated, but each one in a different
manner. Specifically not all the clusters in the sample show the same
dependence of their density profiles on the selected magnitude range, with
NGC 1818 giving evidence of a strong relation and NGC 330 showing only a
hint of the phenomenon. NGC 2004 did not show any significant signature of
mass segregation in its density profiles either. The MFs radial dependence
provides clear proof of the phenomenon for NGC 1818, NGC 2100 and NGC
2004, while for NGC 330 it gives only indications. An investigation of the
constraints introduced by the application of both diagnostic tools is
presented.  We also discuss the problems related to the construction of a
reliable MF for a cluster and their impact on the investigation of the
phenomenon of mass segregation. We find that the MFs of these clusters as
they were constructed with two methods are comparable to Salpeter's IMF. A
discussion is given on the dynamical status of the clusters and a test is
applied on the equipartition among several mass groups in them. Both
showed that the observed mass segregation in the clusters is of primordial
nature.
}

\titlerunning{Mass Segregation in Young MCs Star Clusters with HST}
\authorrunning{D. Gouliermis et al.}
\maketitle

\begin{figure*}[t]
\centerline{\hbox{
\psfig{figure=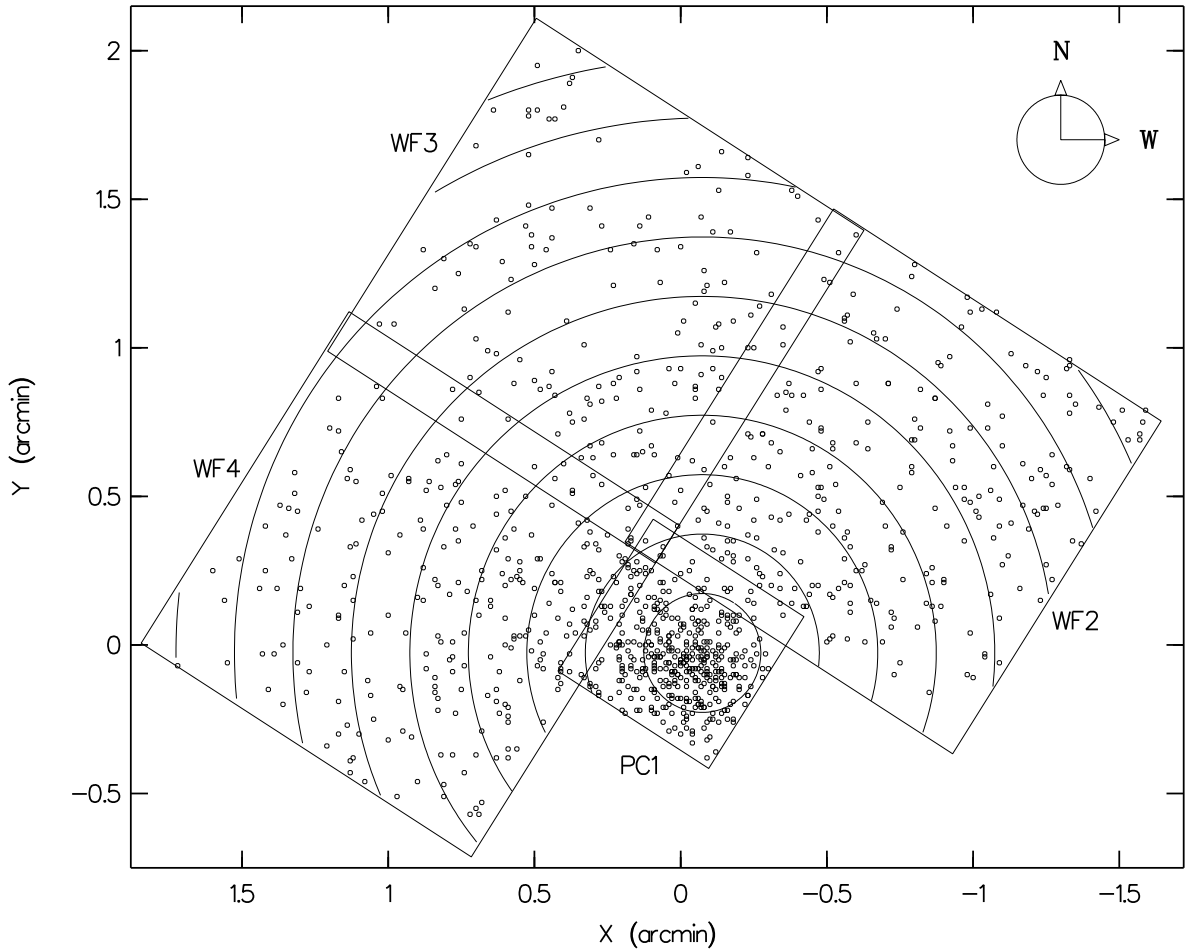,height=13.70truecm,width=19.75truecm,angle=270}
}}
\vspace*{-0.5truecm}
\caption{HST/WFPC2 FOV on NGC 2100. Coordinates are given in arcmin with
(0,0) at the centre of the PC frame. The annuli used for star counts 
have been overplotted (see Sect. 3).}
\label{map}
\end{figure*}

\section{Introduction}

Mass segregation nowadays is a well known phenomenon, which manifests
itself as a strong concentration of the heaviest stars towards the centre
of the cluster. This phenomenon has been extensively observed during the
last decade in a plethora of young open clusters (Jones \& Stauffer 1991;  
Pandey et al. 1992; Shu et al. 1997; Raboud \& Mermilliod 1998;
Hillenbrand \& Hartmann 1998), as well as old globular clusters in the
Galaxy (Cote et al. 1991; Pandey et al. 1992; Paresce et al. 1995; King et
al.  1995; de Marchi \& Paresce 1996; Ferraro et al. 1997; Sosin 1997;
Andreuzzi et al. 2000; Howell et al. 2000). Young star clusters in the
Magellanic Clouds (MCs) have also recently been investigated for mass
segregation (Malumuth \& Heap 1994; Brandl et al. 1996; Fischer et al.
1998;  Kontizas et al. 1998;  Santiago et al. 2001; de Grijs et al.
2002a,b,c; Sirianni et al. 2002). Last, but not least noticeable, is the
work by Stolte et al. (2002), who found mass segregation in the Arches
cluster, a young, massive star cluster and one of the densest in the Milky
Way.

Galactic globulars show mass segregation due to their dynamical evolution,
through which equipartition of kinetic energy among different mass groups
in the cluster can be achieved. According to Spitzer (1969) this leads to
stars of different masses being found in different ``layers'' with the
heavier stars gravitating toward the centre, and thus mass stratification
occurs. This phenomenon is mostly known as {\em Dynamical Mass
Segregation} (for reviews see Lightman \& Shapiro 1978; Meylan \& Heggie
1997). Earlier studies (King 1966; Da Costa 1982; Bolte \& Marleau 1989)
showed that low-mass stars predominate the galactic globulars. This
implies a time-scale for energy equipartition of the order of the
relaxation time of the clusters (the time the system needs to reach a
quasi-Maxwellian equilibrium in its interior). For typical globular
clusters the relaxation time (at the half-mass radius) is significantly
shorter than the cluster age. In consequence, since the cluster had the
time to relax, it exhibits dynamical mass segregation.

The discovery of mass segregation in young clusters (see review by Clarke
et al. 2000) in the MCs introduced new problems concerning its
interpretation, since it can be either the signature of the initial
conditions and loci, where massive stars are formed (e.g. Hillenbrand \&
Hartmann 1998; Bonnell \& Davies 1998), or it can be the result -- at
least partially -- of early dynamics (e.g. Malumuth \& Heap 1994; Brandl
et al.  1996). This phenomenon is usually referred to as {\em Primordial
Mass Segregation}. Observations and theoretical predictions on the
formation of massive stars emphasise the role of early dynamics. An
example has been given by Fischer et al. (1998), who suggest that the
steepening of the mass function they observed at larger radii from the
centre of NGC 2157 is most likely an initial condition of the cluster
formation, since interactions among the protostellar cloudlets play a
crucial role in the massive-star-forming process.

\begin{table*}[t]
\begin{center}
\caption{Ages, metallicities and structural parameters for the sample of
our clusters, as found in the literature. The structural parameters (last
four columns) have been estimated by Mackey \& Gilmore (2003a,b). The core
radii of the clusters are given in col. 6, while $L_{\rm m}$ and $M_{\rm
m}$ are the integrated luminosity and mass to the distance from the centre
of each cluster where a maximum measured extent of its profile is reached
($\simeq 1\farcm3$ for all clusters). The central densities of the
clusters as estimated by Mackey \& Gilmore are given in the last column.}
\begin{tabular*}{\textwidth}[]{@{\extracolsep{\fill}}lcccccccc}
\hline
Cluster & $\log{\tau}$&Age& Metallicity& Met.&$r_{\rm c}$&$\log{L_{\rm m}}$&
$\log{M_{\rm m}}$&$\log{\rho_{\rm 0}}$\\
Name    & (yr)&ref.& [Fe/H]& ref.&(arcmin)&(L{\solar})&(M{\solar})&
(M{\solar} pc$^{-3}$)\\
\hline
\hline
NGC 330 &7.50$^{+0.10}_{-0.50}$& 1,2,3& $-0.82 \pm 0.11$&4  &0.15 $\pm$ 0.01&
5.46$^{+0.12}_{-0.13}$&4.41$^{+0.12}_{-0.13}$&1.84$^{+0.07}_{-0.07}$\\
NGC 2004&7.20$^{+0.30}_{-0.10}$& 1,5  & $-0.56 \pm 0.20$&6  &0.11 $\pm$ 0.01&
5.37$^{+0.16}_{-0.17}$&4.27$^{+0.16}_{-0.17}$&2.32$^{+0.10}_{-0.09}$\\
NGC 1818&7.40$^{+0.30}_{-0.10}$& 1,7    & 0.00 to $-$0.40 &7,8&0.17 $\pm$ 0.01&
5.11$^{+0.10}_{-0.10}$&4.01$^{+0.10}_{-0.10}$&1.55$^{+0.06}_{-0.06}$\\
NGC 2100&7.20$^{+0.20}_{-0.20}$& 1,5  & $-0.32 \pm 0.20$&6  &0.08 $\pm$ 0.01&
5.46$^{+0.22}_{-0.23}$&4.31$^{+0.22}_{-0.23}$&2.64$^{+0.13}_{-0.12}$\\
\hline
\end{tabular*}
\parbox[l]{\textwidth}{References: (1) Keller et al. (2000); 
(2) Chiosi et al. (1995); (3) Da Costa \& Hatzidimitriou (1998); 
(4) Hill (1999); (5) Elson (1991); (6) Jasniewicz \& Th\'{e}venin (1994);
(7) de Grijs et al. (2002b); (8) Johnson et al. (2001)}
\end{center}
\label{params}
\end{table*}

In addition, according to the cluster formation model proposed by Murray
\& Lin (1996) cloudlets fall toward the central region of the
protocluster cloud, and cohesive collisions cause their masses to grow.
More massive stars are subject to many dissipative mergers (Larson 1991)
and so they are preferably formed in the cluster centre (Bonnell \&
Davies 1998), leading to an initial mass segregation. Models suggesting
enhanced accretion rates, which grow with the mass of the accreting
protostar (Behrend \& Maeder 2001), or high collision probabilities in
dense cluster centres (Bonnell et al. 1998) predict the formation of
stars up to 100 M$_{\odot}$ in the densest central regions of a rich
star cluster. Furthermore the total mass and the maximum stellar mass of
a cluster strongly depends on the star formation rate and local density
(Elmegreen 2001), which is enhanced by gravitational collapse or cloud
collisions in high-density environments (Elmegreen 1999, 2001). In
conclusion, protoclusters with high central density are more prone to
exhibit primordial mass segregation.

We investigate the phenomenon of mass segregation in four young star
clusters in the Magellanic Clouds: NGC 1818, NGC 2004 and NGC 2100 in the
Large Magellanic Cloud (LMC) and NGC 330 in the Small Magellanic Cloud
(SMC) with the use of HST/WFPC2 observations (Keller et al. 2000). These
clusters, which are among the younger in the MCs with ages between 10 and
50 Myr, cover a range of metallicities (see Table 1).  This study is based
on a consistent set of HST observations, which allowed us to search for
mass segregation in a uniform manner and to investigate the differences
among the clusters concerning the phenomenon. No such investigation seems
to be available in the literature except for the case presented by de
Grijs et al. (2002a,b,c). In the sample of their clusters these authors
include NGC 1818, while Sirianni et al. (2002) recently presented their
result on the low-mass IMF and mass segregation in NGC 330. The results of
both these studies are based on different data sets than our HST
observations on these clusters. Under these circumstances it would be
interesting to compare our results, which have been preliminarily
presented by Keller et al. (2001) with theirs.

This article is laid out as follows. In Sect. 3 we use our observations to
examine the radial density profiles of the clusters and their dependence
on the selected magnitude range, as an indication of stellar
stratification. Our data were also used in Sect. 4 to derive the mass
functions of the clusters and examine their radial dependence, which is a
widely used method for the detection of mass segregation. The results of
both investigations are given in Sect. 5 and a discussion on the nature of
the observed mass segregation concerning its dynamical origin is presented
in Sect. 6. Finally, Sect. 7 highlights our conclusions.

\begin{figure}
\centerline{\hbox{
\psfig{figure=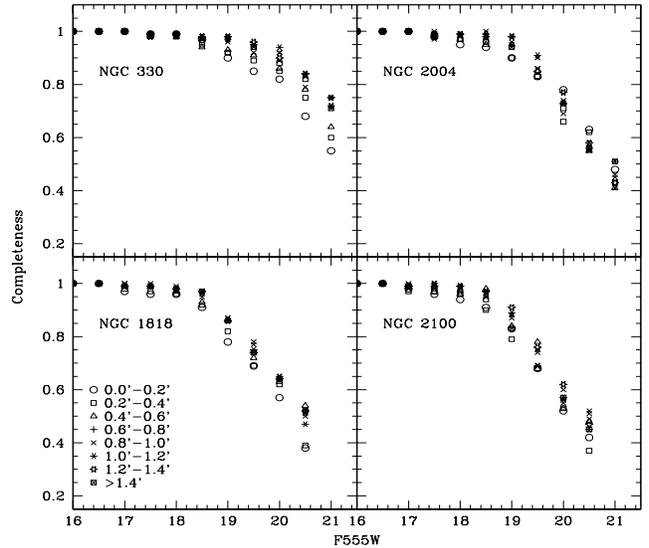,height=8.0truecm,width=9.truecm,angle=0}
}}
\caption{Completeness of the WFPC2/F555W photometry for 
various radial distances for the observed star clusters.}
\label{compl}
\end{figure}

\section{Observations and Data Reduction}

The observations of the clusters have been discussed in Keller et al.
(2000), where more details on the photometric reductions are also given,
together with the magnitudes and errors for each frame and filter.  
The raw frames were bias-subtracted and flat-fielded in the usual way
using the standard WFPC2 pipeline. For the present study we consider the
star counts and luminosity functions derived from the $F555W$ ($\equiv
V$). In Sect. 4.2 we also use the $F160BW$ ($\equiv$ far-$UV$) to
examine the population on the upper main sequence. An example of the
HST/WFPC2 field of view (FOV) of these observations is given in Fig.
\ref{map} for NGC 2100.

\subsection{Completeness Corrections}

We establish a quantitative estimation of the degree of completeness
through extensive artificial star tests. A set of 100 artificial stars was
generated within each of the four frames of the WFPC2 images for every
cluster with a range of magnitudes. In addition, the distribution of
objects on the original frame was used to determine the placement of
artificial stars within the frame. This was repeated 10 times. A 
reduction procedure was performed on the enriched frames identical to that
performed on the original frames. An artificial star was considered as
``recovered'' if the recovered image centroid agrees with the actual
position to within 1 pixel and if the recovered magnitude is within 0.2
mag of the actual magnitude. The completeness factor is the ratio between
the number of artificial stars recovered to the number of stars originally
simulated.

Using the above procedure, we have measured the completeness within our
F555W frames for various radial distances from the centre of each cluster.
Completeness factors are shown in Fig. \ref{compl}, in which one can see
that completeness is a moderate function of radius but mostly a function
of magnitude. This is probably because the cluster cores are not
excessively crowded at the scale of the WFPC2. We use these factors for
the completeness corrections applied to the observational luminosity and
mass functions and star counts. It should be noted that only data for
which the completeness is better than 70\% were actually used in the 
analysis that follows.

\begin{figure} 
\centerline{\hbox{
\psfig{figure=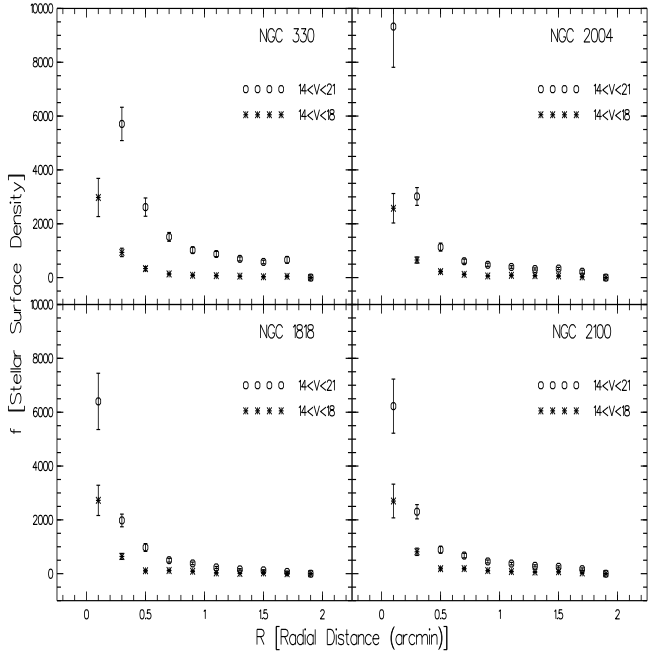,width=9.15truecm,height=9.15truecm,angle=270} 
}}
\caption{Radial distribution of stellar density, $f$, of all four 
clusters for stars in two magnitude groups.}
\label{profraw} 
\end{figure}

\begin{figure*}
\centerline{\hbox{
\psfig{figure=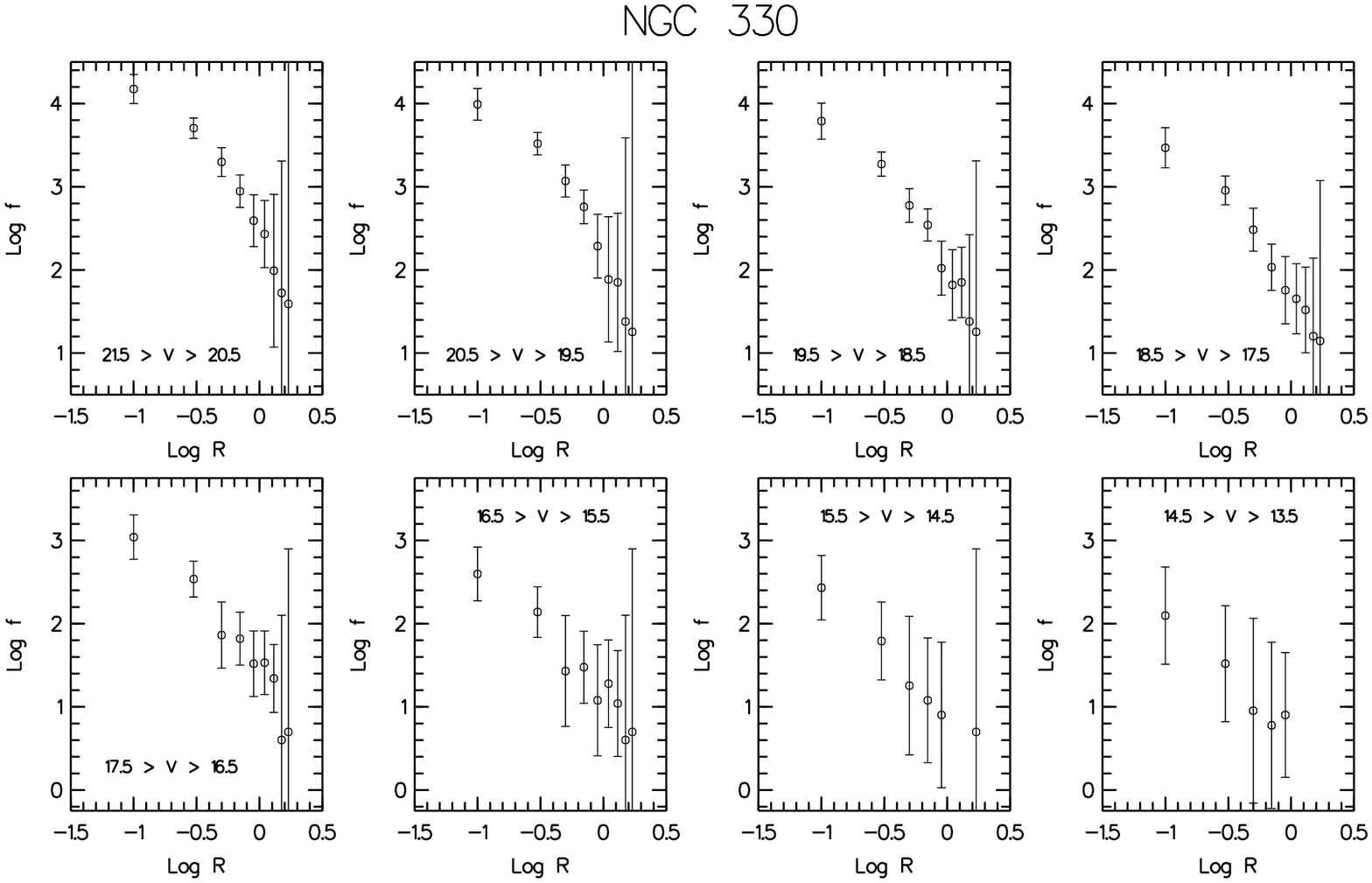,height=8.25truecm,width=14.truecm,angle=270}
}}
\vspace*{-1.05cm}
\centerline{\hbox{
\psfig{figure=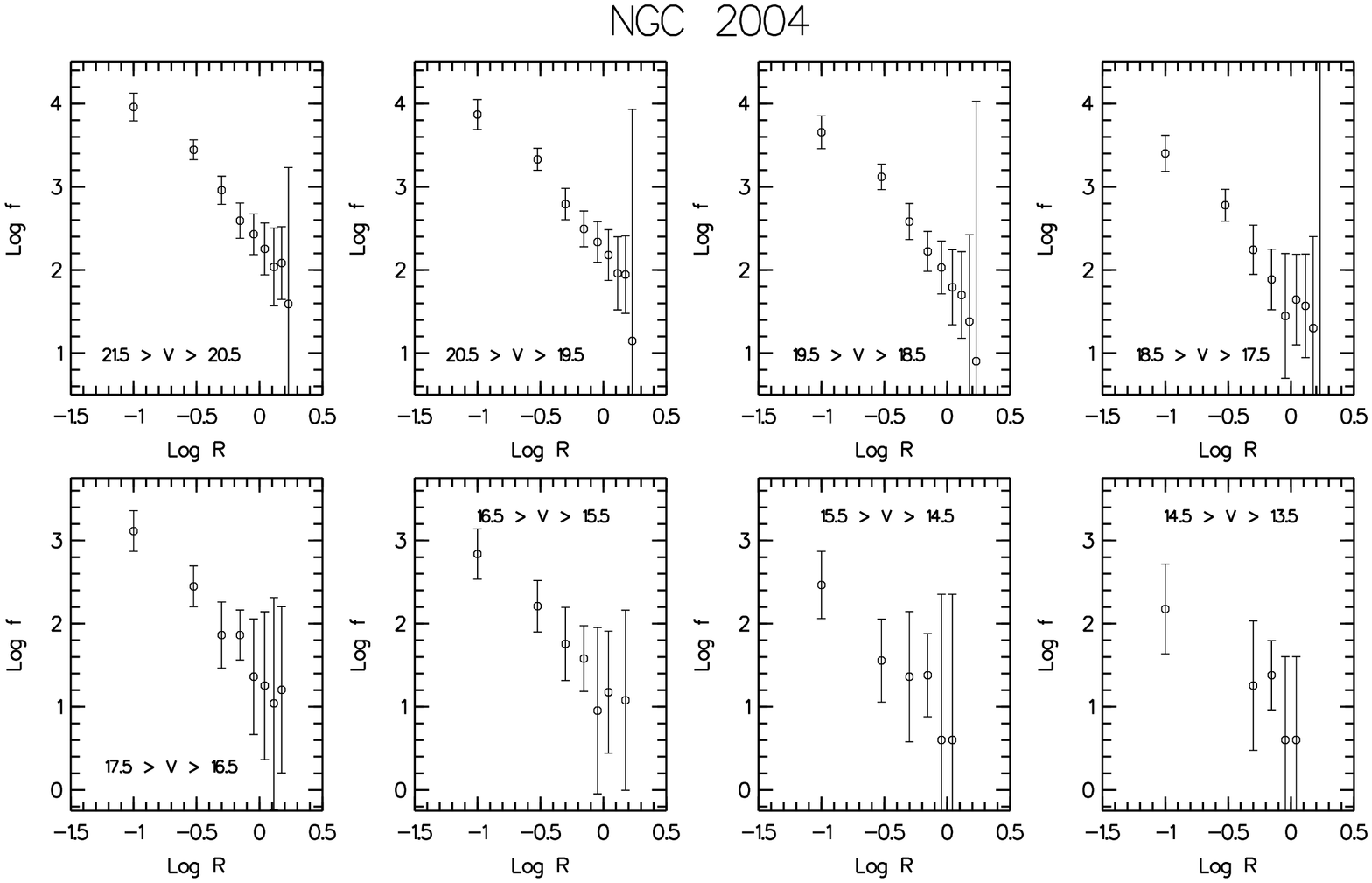,height=8.25truecm,width=14.truecm,angle=270}
}}
\caption{Surface density profiles of NGC 330 and NGC 2004 for stars in 
various magnitude ranges.}
\label{profs}
\end{figure*}

\setcounter{figure}{3}
\begin{figure*}
\centerline{\hbox{
\psfig{figure=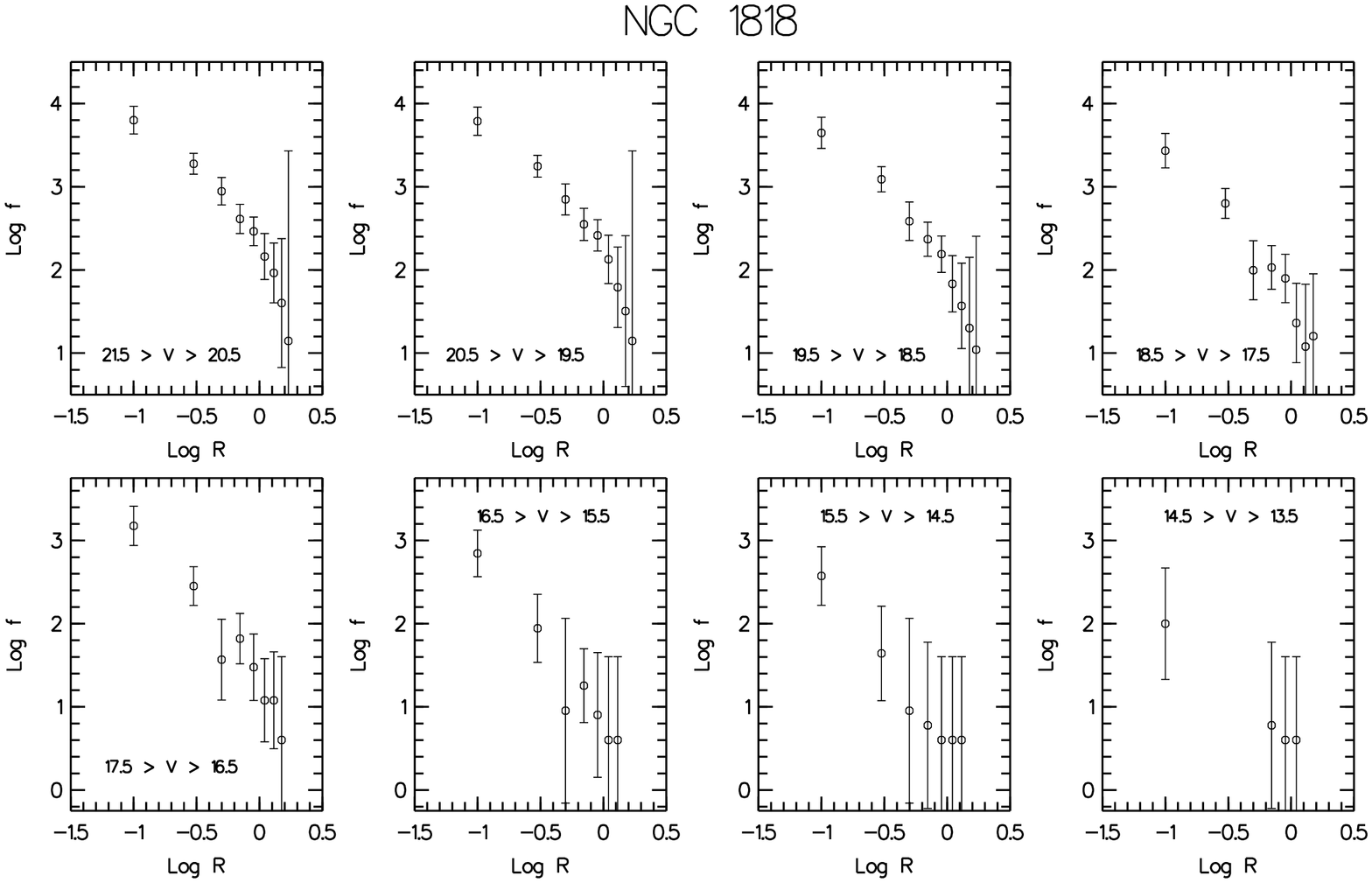,height=8.25truecm,width=14.truecm,angle=270}
}}
\vspace*{-1.05cm}
\centerline{\hbox{
\psfig{figure=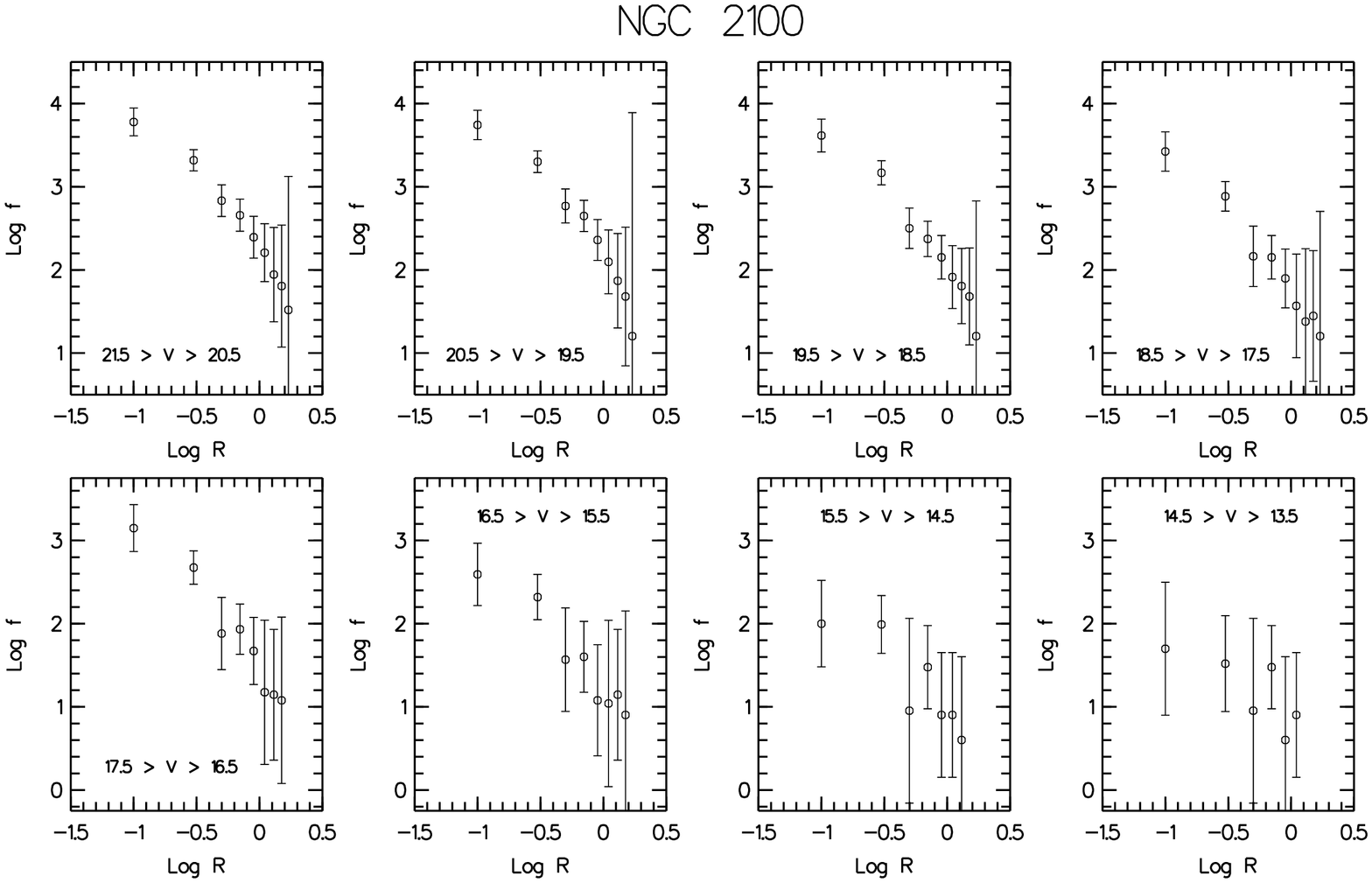,height=8.25truecm,width=14.truecm,angle=270}
}}
\caption{{\em (Continued).} Surface density profiles of NGC 1818 and NGC 
2100 for stars in various magnitude ranges.}
\label{profs}
\end{figure*}

\section{Surface Density Profiles of the Clusters}

To investigate the variation in the radial density profiles of the
clusters for stars of different magnitudes we have performed star counts
at selected radial distances and several magnitude slices for each
cluster.  The surface stellar density $f$ is derived as the number of
stars per unit area corrected for i) incompleteness as defined in the
previous section, and ii) for the field contribution, as shown in 
the following section.

\subsection{Field Star Contamination}

Our observations did not include offset field frames from the purpose of
field star subtraction. Thus, in order to estimate the contribution of
field stellar density (background density) to the profiles of the clusters
we plotted the number of counted stars per unit area (in arcmin$^{2}$), 
$f=N/A$, corrected for incompleteness at selected radial rings.  As is
expected, the surface stellar density, $f$, drops as a function of 
distance, $R$, from the centre of the cluster. This is shown in Fig.  
\ref{profraw} of the profiles of all four clusters for stars in two 
selected magnitude ranges. The density numbers are normalised to the same 
surface area.

In Fig. \ref{profraw} it is shown that for all the clusters beyond a
radius $R$ \gapprox\ 1\farcm5, the number of stars per unit area drops to
a uniform level, which might be considered as a good approximation of the
background density.  Considering that the WFPC2 FOV on our clusters is not
wide enough to check if the density profiles are still flat further away
than 1\farcm5, one should expect the value of the background density to be
overestimated.  We checked the validity of our determination for the field
contribution on NGC 1818, for which offset field data exist (Santiago et
al. 2001), which we reduced in the same manner. In each magnitude
range considered in Fig. \ref{profraw} we found, within the uncertainties,
the same number of field stars with the one found for radial distance $R
>$ 1\farcm5 in our frames. Furthermore, Vallenari et al. (1994) found that
the radius of NGC 330 is about 95\arcsec and Sirianni et al. (2002)
assumed that the measured stars in the most external region of their WFPC2
FOV beyond this distance from the centre of the cluster represent an
estimate of the background field. This radial distance of $\sim$ 1\farcm58
is in good agreement with our selected limit of the cluster.

It should be noted that the background density, shown in Fig.
\ref{profraw} for distances larger than $R \simeq$ 1\farcm5 from the
centre of each cluster, does not seem to differ significantly from one
cluster to the other, being very small and representing for both selected
magnitude ranges about 1 to 2\% of the total population. Mackey \& Gilmore
(2003a,b) report that their detailed surface profiles of these clusters
derived from HST observations, seem to have a maximum extent of
72{\arcsec} for NGC 330 and 76\arcsec\ for the other three LMC clusters.
The two measurements are very close to each other and well below our
selected radial limit of 1\farcm5. All the above gives confidence that we
have not significantly overestimated the field population with this
technique. Having so defined the contribution of the background ($b$)  we
subtract it from the completeness corrected star counts per unit area
($N/A$) to estimate the corresponding surface stellar density defined now
as $f=N/A-b$. All numbers are normalised to the same surface area.

\subsection{Density profiles}

The surface density profiles were derived by counting stars in 8 different
magnitude bins (from $V$ around 14 to 21 mag) from the F555W observations.
The counts were performed in 9 radial annuli around the centres of the
clusters, spanning 0\farcm2 each. The plots of $\log{(f)}$ versus
$\log{(R)}$ for the selected magnitude bins, which almost cover the
observed magnitude range (down to the detection limit), are shown in Fig.
\ref{profs}.  Taking into account that the density profiles can be
approximated by $\log{(f)} \propto \gamma \times \log{(R)}$ (Elson et al.
1987), the slope $\gamma$ is a very good indication of the existence of
mass stratification at the magnitude where this slope changes
significantly (e.g. Subramaniam et al.  1993). In order to investigate
such a relation we estimated the slope $\gamma$ of the density profile in
each magnitude range by using the data of the rings in which stars were
counted except the two outer rings of distances around 1\farcm5 and
1\farcm7 from the cluster's centre each. This was done due to low number
statistics at these distances, since the field contribution at these
distances is severe.

\begin{figure*}
\centerline{\hbox{
\psfig{figure=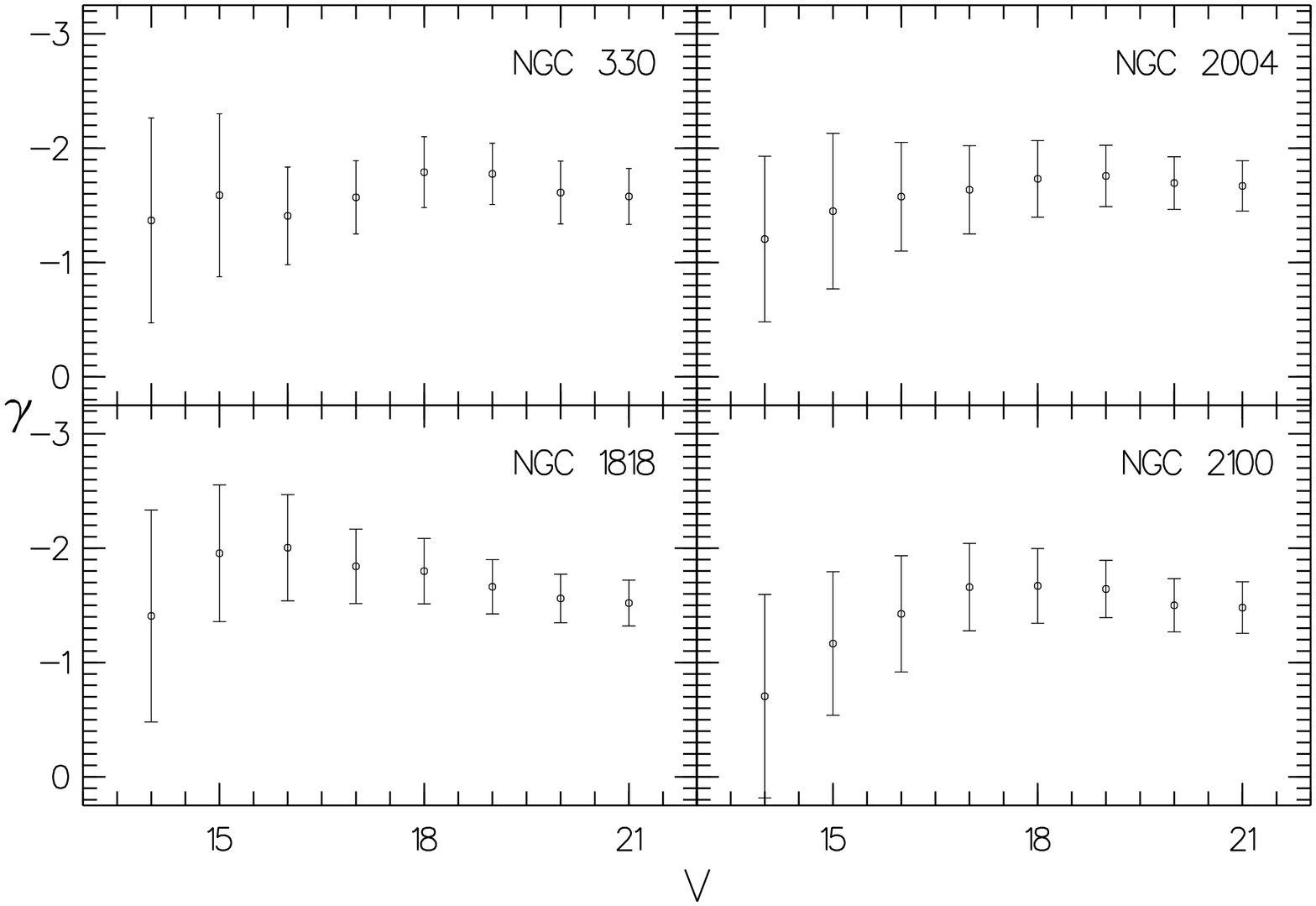,width=13.75truecm,angle=270}
}}
\caption{Relation of the slope $\gamma$ of the density profiles to
the corresponding magnitude range for the clusters of our sample.}
\label{denslp}
\end{figure*}

The correlation of the estimated density profile slopes $\gamma$ with 
the corresponding magnitude bin is shown in Fig.  
\ref{denslp}. The two or three first points of these diagrams, which
cover a magnitude range of $V \approx$ 14 - 16 mag can be questioned,
because of their large errors. For the points corresponding to fainter
magnitudes one may see that there are cases where $\gamma$ actually
depends on the magnitude bin. Such is the case of NGC 1818, which implies  
that mass stratification is actually taking place in the cluster. The 
relations $\gamma$ versus magnitude shown in Fig. \ref{denslp} were 
linearly fitted using the least squares solution for every cluster and the 
corresponding Spearman correlation coefficients were estimated. This was 
done using the points in almost all the magnitude ranges (from $V=$ 15 - 
21 mag down to $V=$ 18 - 21 mag) in order to specify the magnitude 
range where any significant correlation can be found. The results of 
these statistical tests are presented in Table 2, where we give the 
estimated slope of the linear fits and the corresponding probability that 
the data are indeed correlated, as was estimated from Spearman's method 
for all the magnitude ranges used.

From the values of Table 2 it is clear that indeed there is a strong
correlation of $\gamma$ with magnitude for almost all the magnitude ranges
in NGC 1818 (with a slope $\sim$ 0.1). For NGC 330 and NGC 2100 this
correlation becomes significant for $V\geq18$ mag (with slopes 0.08 and
0.07 respectively). In the case of NGC 2100 a significant correlation
(with 90\% probability) can be considered also for the magnitude range $V
\simeq$ 17 - 21 mag. Finally for NGC 2004 it is found that there is a weak
dependency between $\gamma$ and magnitude limit, since the most important
slope found (for V \gapprox\ 18 mag with 80\% correlation probability) is
very shallow. In consequence, NGC 1818 is expected to exhibit the
phenomenon strongly, while NGC 330 and NGC 2100 can be seriously
considered as candidates for mass segregation. NGC 2004 according to its
surface density profiles cannot be considered as an important candidate.

\begin{table}
\begin{center}
\caption{Linear fit slopes and corresponding correlation probabilities
(P) of $\gamma$ versus magnitude for various selected magnitude ranges, 
as shown in Fig. \ref{denslp}, for every cluster in our sample.}
\begin{tabular}{c|cr|cr}
\hline
mag&linear fit&P&linear fit&P\\
range& slope& (\%)& slope& (\%)\\
\hline
&\multicolumn{2}{c}{NGC 330}&\multicolumn{2}{c}{NGC 2004}\\
\hline
15 - 21& $-0.021 \pm 0.025$&  32& $-0.036 \pm 0.014$&  68\\
16 - 21& $-0.027 \pm 0.036$&  43& $-0.019 \pm 0.015$&  49\\ 
17 - 21& $~~0.016 \pm 0.018$&   0& $-0.003 \pm 0.017$&  10\\ 
18 - 21& $~~0.080 \pm 0.020$& 100& $~~0.025 \pm 0.012$&  80\\ 
\hline
&\multicolumn{2}{c}{NGC 1818}&\multicolumn{2}{c}{NGC 2100}\\
\hline
15 - 21& $~~0.085 \pm 0.009$&  96& $-0.038 \pm 0.033$&  32\\
16 - 21& $~~0.097 \pm 0.008$& 100& $~~0.007 \pm 0.028$&   9\\
17 - 21& $~~0.088 \pm 0.010$& 100& $~~0.053 \pm 0.014$&  90\\
18 - 21& $~~0.094 \pm 0.015$& 100& $~~0.071 \pm 0.017$& 100\\
\hline
\hline
\end{tabular}
\end{center}
\label{gamma}
\end{table}

Subramaniam et al. (1993) investigated the magnitude dependence of the
slope of the density profiles in the halo regions of five young LMC
clusters, NGC 2004 and NGC 2100 being among them. These authors did not
find any significant correlation of that kind for these two clusters
in their CCD photometry from the 1.54m Danish telescope at ESO. It is
worth noting that with our higher resolution observations we were able
to detect only a weak correlation of $\gamma$ and magnitude range for
these clusters, as an indication that they are probably segregated.

In conclusion, the surface density profiles and their dependency on the
magnitude bins of counted stars provides the first evidence of mass
segregation in clusters, giving an estimation of the magnitude (mass)
limits for the segregated stars. NGC 330 and NGC 2004 were found to have
segregated stars for $V \simeq$ 18 mag. The limit for NGC 2100 is $\sim$
17 mag, while for NGC 1818 is 15 - 16 mag. From the results presented
above it seems that the clusters exhibit a different degree of mass
segregation as far as the dependence of the density profiles on the
selected magnitude ranges is concerned.

In any case this diagnostic tool cannot be considered of high accuracy,
due to the large uncertainties in the estimated slopes of the density
profiles of the clusters and it is not usually applied in the recent works
on mass segregation. We present it here to show that the problem of low
accuracy, which was known for ground-based CCD observations (e.g. Kontizas
et al. 1998), is also apparent in the case of HST data. On the other hand,
a more accurate diagnostic tool for the detection of mass segregation is
the analysis of the luminosity and mass function (MF) of the cluster, and
it is widely used. Practically it is the only tool available which makes
use of photometric observations. We present, thus, our results on its
application to our clusters in the following section.

\begin{figure*}
\centerline{\hbox{
\psfig{figure=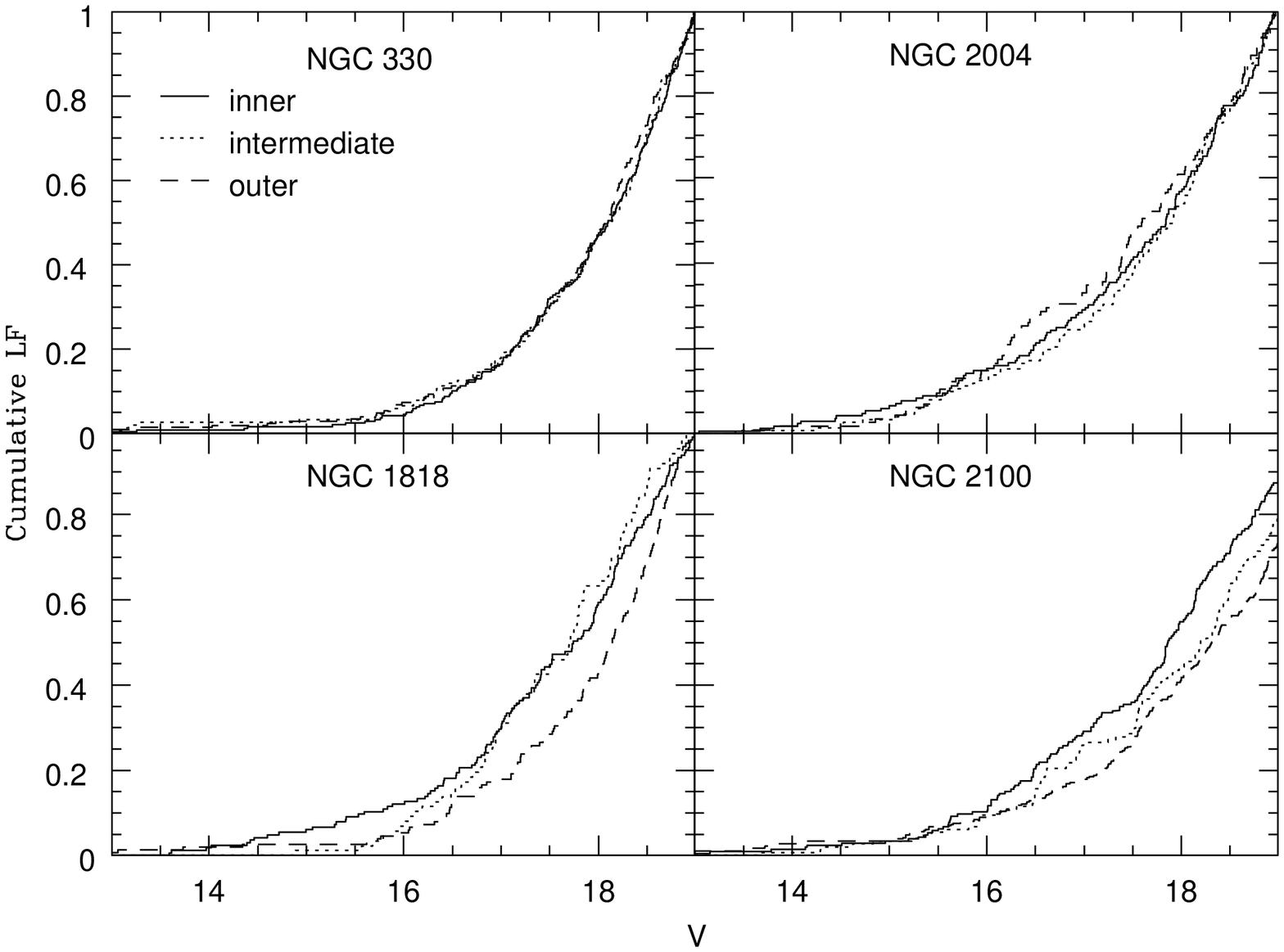,height=9.6truecm}
}}
\caption{The cumulative luminosity function for stars on the main-sequence
in the clusters of our sample in three radial zones.}
\label{lfs}
\end{figure*}

\section{Luminosity Functions and Mass Functions of the Clusters}

According to this diagnostic, if there is a higher concentration of
massive stars toward the centre of the cluster, then mass segregation
should be exhibited from a radial variation of its MF slope: The radial
MFs of the cluster should become steeper outwards, where less high-mass
stars are expected to exist. In consequence, at a certain distance the
change of the slope of its mass (or luminosity) function becomes evident
if mass segregation occurs. Before we proceed to the construction of the
MFs of the clusters we use their LFs and we present a preliminary
investigation on their radial dependence (if any) at selected radial
distances from the centre of the clusters.

\subsection{Luminosity Functions}

For the investigation on the LFs of the clusters we focus on the
main-sequence (MS) populations of the clusters within three regions:  
inner ($r <0.\arcmin2$), intermediate ($0.\arcmin2 < r < 0.\arcmin4$) and
outer ($r>0.\arcmin4$). To improve the number statistics we use the
normalised cumulative LFs. We have corrected our sample for radially
variable completeness and field contamination as before. Fig. \ref{lfs}
illustrates the cumulative LFs for the three regions in every cluster. The
data set has been truncated at $V=19$ mag at which point the completeness
is nowhere less than 80\% and field stars contribute no more than 10\% to
the uncorrected sample. In Fig. \ref{lfs} a radial variation of the LFs of
some of the clusters is apparent for the three selected annuli. In order
to quantify these variations we applied the Kolmogorov-Smirnov (KS) test
on the data shown in the figure. This statistical test indicates the
probability that two distributions are derived from an identical parent
distribution, thus it offers a pretty clear indication of the significance
of the differences between the LFs at different distances from the centre
of each cluster.

We applied the KS test in order to compare the LFs of the clusters in the
inner annulus ($r <0.\arcmin2$) with those in the intermediate
($0.\arcmin2 < r < 0.\arcmin4$) and outer ($r>0.\arcmin4$) annulus. The
test showed that the probability of common origin for the LFs in the inner
and intermediate regions of NGC 1818 and NGC 2100 is only 2\% and 8\%
respectively.  The difference between the LFs in the inner and outer
regions of these clusters is even more significant, having a probability
of common origin of 1\% for NGC 1818 and 0.3\% for NGC 2100. The
probabilities for NGC 330 and NGC 2004 are totally different. We found
that the probability that the LFs of the inner and intermediate regions of
NGC 330 have common origin is 69\%. This probability for the LFs of the
inner and outer regions of NGC 330 is 76\%. These numbers are even higher
for NGC 2004, equal to 98\% for the inner-intermediate LFs and 78\% for
the inner-outer LFs of the cluster.

These results suggest that the LFs of the clusters NGC 1818 and NGC 2100
give clear evidence of the phenomenon of mass segregation, while NGC 330
and NGC 2004 show only indications of it. It should be noted though, that
the LFs of the clusters were used here only as indicative diagnostic for
any luminosity stratification in {\em selected radii} around the centres
of the clusters. In the following section we construct the MFs and
investigate the radial variation of their slopes systematically for the
clusters in a more thorough manner. We use our results on the MFs along
with the ones of this section on the LFs in order to demonstrate the
sensitivity of any conclusion concerning mass segregation to the selected
annuli around the centre of the cluster.

\subsection{Mass Functions}

The distribution of stellar masses formed in a given volume of space in
a stellar system, known as Mass Function, can be represented by the
Initial Mass Function (IMF) assuming that all stars in the system are
the product of a single star formation event. This approximation is valid
for young star clusters in the MCs, such as the clusters of our sample.
There are various parametrisations of the IMF (see e.g. Kroupa 
2002). A widely used one was proposed by Scalo (1986), where the 
IMF is characterised by the logarithmic derivative $\Gamma$, called index:
\beq 
\Gamma = \frac{{\rm d}\log{\xi(\log{\rm m})}}{{\rm d}\log{\rm m}} 
\eeq
where $\xi(\log{\rm m})$ is the IMF. $\Gamma$ is its slope and can be
derived from the linear relation of $\xi(\log{\rm m})$ and
$\log{\rm m}$. It is a common practice to use as a reference value the
index $\Gamma$ as found by Salpeter (1955) for the solar neighbourhood
($\Gamma = -1.35$, for a mass range 0.4 \lapprox\ $m$/M{\solar} \lapprox\
10). The MF of a system is constructed by counting stars in mass
intervals. This can be achieved by two methods: The first method by 
directly counting stars between evolutionary tracks according to their 
positions in the HRD (e.g. Massey et al. 1995) and the second method by 
translating their luminosities into masses using mass-luminosity relations 
(e.g. de Grijs et al. 2002b) and then constructing the distribution of the 
derived masses.

From here on we will refer to these methods as $first$ and $second$
counting $method$ respectively. It is certain that the $first$ $method$ is
more accurate concerning the mass bins where the observed stars are going
to be distributed. This is because no direct transformation of
luminosities to masses is required, but the construction of the MF is more
straightforward, since it is made by counting stars according to their
observed absolute magnitudes and colours (corrected for reddening). For
the $second$ $method$ there is a definite dependence on the isochrone
models used for the transformation of luminosities to masses. De Grijs et
al. (2002b) argue that the conversion of an observed LF to its associated
MF is not as straightforward as is often assumed. Thus in order to achieve
a reliable conversion, one needs to have accurate knowledge of the
appropriate mass-luminosity relation.

In addition, problems occur if there is an age distribution among the
stars in the cluster and so the use of a single isochrone may not be
adequate for a mass-luminosity conversion. De Grijs et al. note that small
differences between MF slopes as a function of age are appreciable for
younger stellar populations. Stolte et al. (2002) found that the use of
isochrones for older stellar populations will result in flatter MFs.  
Still, the main-sequence mass-luminosity relation from isochrones of
different age should be almost the same. What would be of concern is how
strong the impact of different turn-off points would be on the MF slope.
In any case, since these differences are systematic, they should not be
expected to alter any radial dependence of the cluster MF.

As far as the $first$ $method$ is concerned, there is also a dependence on
the theoretical models, but one achieves smaller uncertainty by using
information on both magnitudes and colours. Furthermore, the use of
evolutionary tracks, between which the stars can be directly counted on
the HRD becomes even more straightforward if one is interested only in the
MS stars (as is the case here), since the evolutionary tracks on the MS
are more or less parallel to each other for stars of different masses, and
so finding the distribution of stars in mass bins is rather simple. Still,
there is a constraint in applying this method, since one has to use only
the stars detected in both wavebands, and thus a smaller stellar sample
for the construction of the MF. The problem becomes important due to the
filters of our HST observations, $F555W$ and $F160BW$, which correspond to
standard $V$ and far-$UV$ respectively, because our short exposures give a
small sample of stars found in both filters and they allow us to report
results only on the upper main-sequence MFs of the clusters.

In consequence, in order to compare our results on the MF with different
methods, and to have a better statistical significance for the
investigation of the phenomenon of mass segregation in our clusters we
also applied the $second$ $method$ for the construction of their overall
and radial MFs. For the $first$ $method$ we used the evolutionary tracks
by Schaerer et al.  (1993) for the metallicity of the LMC clusters
(Z=0.008) and by Charbonnel et al. (1993) for NGC 330 (Z=0.004). In both
sets of grids mass loss and moderate core overshooting are taken into
account. We compared these evolutionary tracks on the HRD with the
isochrones used by Keller et al. (2000) from Bertelli et al.  (1994) for
the estimation of the ages of the clusters and we found an excellent
agreement on the main sequence.

For the $second$ $method$, the LFs of the clusters were used to determine
the MFs adopting a mass-luminosity relation derived from the used
theoretical models. In both cases we selected the MS stars, which were
counted in logarithmic (base ten) mass intervals. The counted numbers were
corrected for incompleteness and were field-subtracted. We assumed, as in
Sect. 3, that the field population is well represented by the stars with
radial distances larger than about 1\farcm5.  Finally, the counted numbers
were normalised to a surface of 1 kpc$^{2}$. The errors in the derived MFs
reflect the Poisson statistics of the counting process and they are
suitably corrected and normalised.

\begin{figure*}
\centerline{\hbox{
\psfig{figure=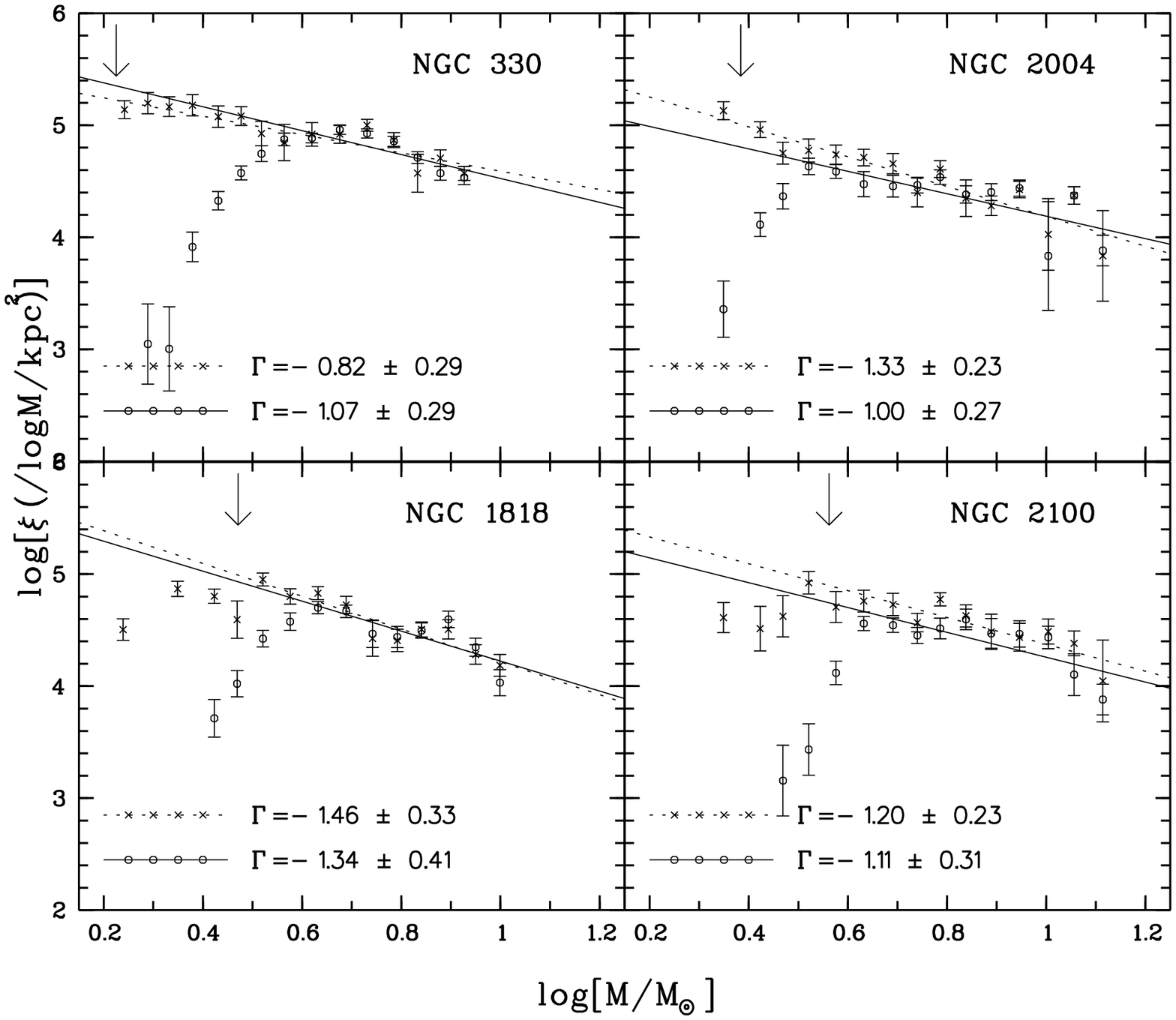,width=15.35truecm,angle=270}
}}
\vspace*{-0.75truecm}
\caption{Mass Functions of the MS stars of the clusters. They were
constructed with two methods (see Sect. 4.2): Counting stars between
evolutionary tracks on the HRD ($\circ$) and using a mass-luminosity
relation based on the theoretical models used for the determination of the
ages of the clusters ($\times$).  The MF index $\Gamma$ is given as
estimated in both cases for the same mass range for reasons of comparison.
The corresponding slopes have been overplotted with solid lines for the MF
constructed with the $first$ $method$ and with dashed lines for the
$second$. The derived MF slopes do not differ significantly, while their
differences toward the low-mass end are purely statistical due to the
larger numbers of stars used for the $second$ $method$, since for the
$first$ $method$ we could only use stars found with both $F555W$ and
$F160BW$ filters. The indices of the MFs were found adopting a
single-power law and they are given for the most complete mass ranges
available ($\sim$ 3 - 9 M{\solar} for NGC 330, 3 - 14 M{\solar} for NGC
2004, 4 - 11 M{\solar} for NGC 1818 and 4 - 14 M{\solar} for NGC 2100).
Arrows indicate the limit of 70\% of completeness.}
\label{mfsfig}
\end{figure*}

We compared the MF slopes derived with both methods, and it was found that
they do not differ considerably, as is shown in Fig. \ref{mfsfig}, where
we present the MFs of the clusters throughout the observed
areas (overall MFs). The differences shown in the MFs toward the low-mass
end are due to the smaller stellar sample with available magnitudes in
both {\em F555W} and {\em F160BW} used for the $first$ $method$. We also
checked the MF slopes derived with both methods for selected annuli around
the centres of the clusters and we found that these differences are
becoming more significant, possibly because of fewer stars counted within
the annuli in both cases. Still, if there was a radial dependence of
$\Gamma$, it was apparent for the MFs constructed with both methods. Thus
the two counting methods seem equally adequate for the investigation of
mass segregation. For the study of the radial MFs of the clusters, though,
we used the slopes derived with the $second$ $method$, because of better
statistics.

\begin{figure*}
\centerline{\hbox{
\psfig{figure=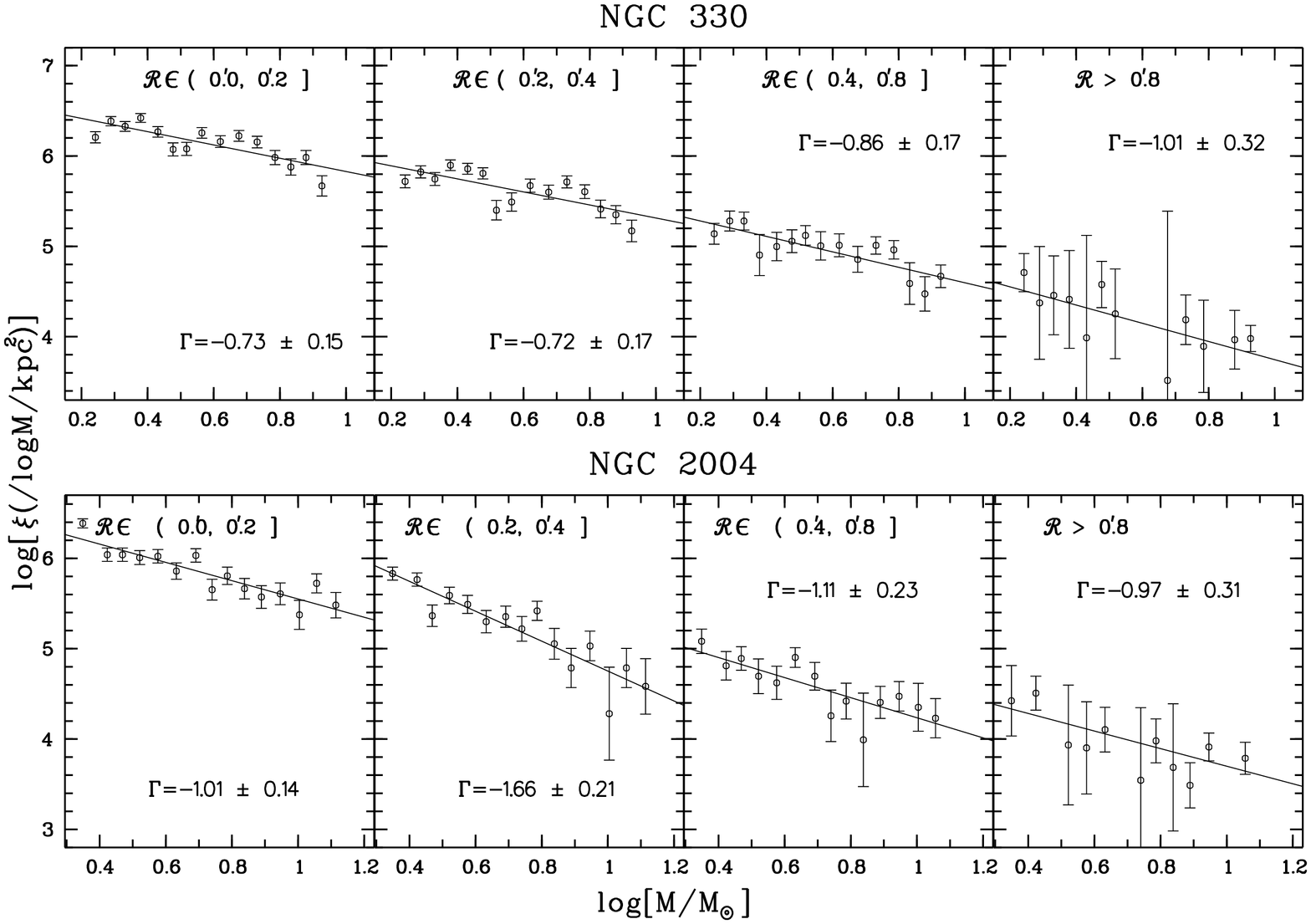,width=15.75truecm,angle=270}
}}
\vspace*{-0.35truecm}
\caption{Mass Functions of the clusters NGC 330 and NGC 2004 for selected
radial distances: 0\farcm0 - 0\farcm2, 0\farcm2 - 0\farcm4, 0\farcm4 -
0\farcm8 and for distances larger than 0\farcm8.  From the selected annuli
it is shown that the MF of NGC 330 seems to remain more or less unchanged,
while that of NGC 2004 becomes flat outwards. Both MFs become very noisy
outwards, due to the fact that the clusters are sparse at radii larger
than 0\farcm8. The slopes drawn are estimated for every cluster in the
same mass ranges as those of the overall MFs given in Fig. \ref{mfsfig}.}
\label{rdmffig}
\end{figure*}

\setcounter{figure}{7} 
\begin{figure*} 
\centerline{\hbox{
\psfig{figure=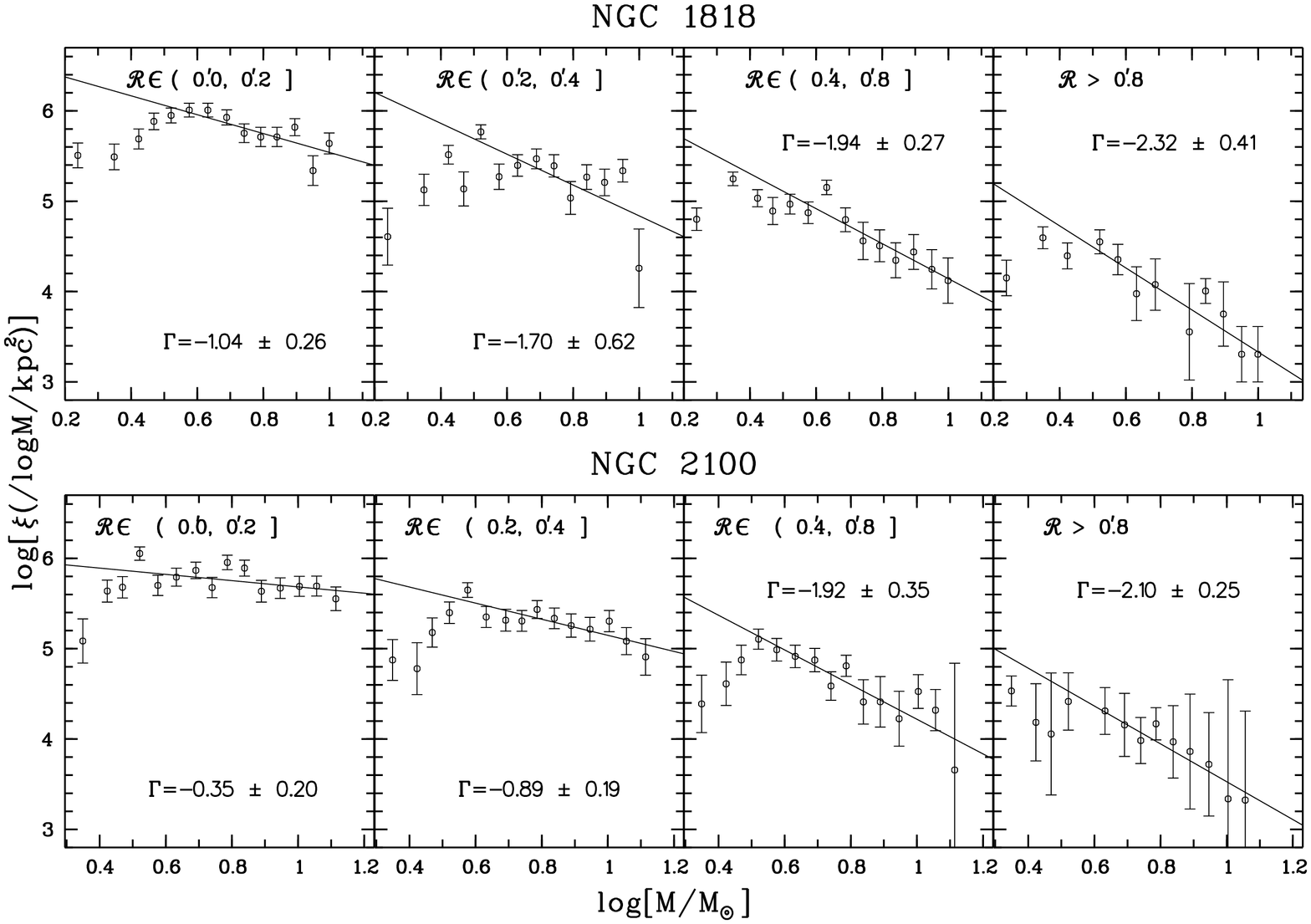,width=15.75truecm,angle=270} 
}}
\vspace*{-0.35truecm}
\caption{{\em (Continued)}. Mass Functions of the clusters NGC 1818 and
NGC 2100 for indicative selected radial distances: 0\farcm0 - 0\farcm2,
0\farcm2 - 0\farcm4, 0\farcm4 - 0\farcm8 and for distances larger than
0\farcm8. The dependence of the MF slope to the selected annuli is rather
obvious for these clusters. The drawn slopes are estimated for every
cluster in the same mass ranges as those of the overall MFs given in Fig.
\ref{mfsfig}.}
\label{rdmffig}
\end{figure*}

\subsubsection{Mass Functions of the clusters}

The overall MFs of the clusters are shown in Fig.  \ref{mfsfig}. These MFs
give an upper mass limit, which is similar for both counting methods.
Massive stars up to $\sim$ 14 M{\solar} were found in NGC 2004 and NGC
2100, while for NGC 330 and NGC 1818 the largest observed masses are
$\sim$ 9 M{\solar} and $\sim$ 11 M{\solar} respectively. As far as the
low-mass end is concerned, though the MFs are corrected for incompleteness
we could not use the entire mass range for the estimation of their slopes
due to bad statistics toward the lower mass bins, because of the high
detection limit of our observations. This ``incompleteness'' toward the
low-mass end seems to be connected to the total observed populations of
the clusters and to the stellar samples used in every counting method for
the derivation of the MFs. For example the most complete samples are the
ones for NGC 330 and NGC 2004, which are the most populous clusters in the
sample. We therefore calculated the MF slopes by selecting mass ranges
that comprise stars with masses down to the most complete limit for both
methods used. This mass limit corresponds to $\sim 3$ M{\solar} for NGC
330 and NGC 2004 and to $\sim 4$ M{\solar} for NGC 1818 and NGC 2100.

Fig. \ref{mfsfig} shows that the MFs may not be single-power law
distributions. This is more obvious for NGC 330 and NGC 1818, where it
seems that the overall MFs of these clusters can be separated in at least
two mass groups following different trends. Still, the use of a
single-power law (linear fit) for an estimation of the MF index is a
common practice, which we also follow for two reasons: (1) to be able to
compare our results with previously published ones and (2) to identify
more easily any radial dependence of the MFs. The MF slopes as calculated
using the least-square linear method for the selected mass ranges are
overplotted in Fig. \ref{mfsfig} with solid lines for the MFs constructed
with the $first$ $method$ and with dashed lines for those constructed with
the $second$.  The fits are not weighted. From the estimated slopes, which
are also given in Fig. \ref{mfsfig}, one can see that NGC 1818 exhibits
the steeper MF, while NGC 330 is seen to have the flattest MF in the
sample. If we consider the slopes as found with either method for the MF
construction and their uncertainties, all clusters seem to have MF slopes
around Salpeter's value, with those of NGC 1818 and NGC 2004 being closer
to it.

\subsubsection{Radial Mass Functions of the clusters}

In order to check any radial dependence of the MF slope, as indication of
mass segregation, we constructed the MFs of the clusters in several
selected annuli around the centre of the clusters. For this investigation
we use the slopes of the radial MFs, as they were constructed only with
the $second$ $method$, since the $first$ $method$ was found inadequate for
the construction of the radial MFs due to poor stellar numbers. For the
inner area of each cluster we selected four annuli spanning 0\farcm1 each,
while for distances larger than 0\farcm4 we used annuli 0\farcm2 wide. To
check for any variability in the derived radial dependence of the
estimated slopes for different selected radial distances, we estimated the
MF slopes of the clusters also in thinner annuli. It was found that the
observed MF radial dependence is rather sensitive to the radial distances
of the annuli selected. This can also be demonstrated by the LFs of NGC
330 and NGC 2004 (Sect. 4.1), since for three selected annuli around their
centres no mass segregation was observed, while the shape of their MFs in
different annuli is totally different, especially for NGC 2004, as is
shown in detail below.  Thus, one may conclude that a unique set of
selected annuli with widths increasing outwards cannot be chosen as easily
as it seems.

\begin{figure*}
\centerline{\hbox{
\psfig{figure= 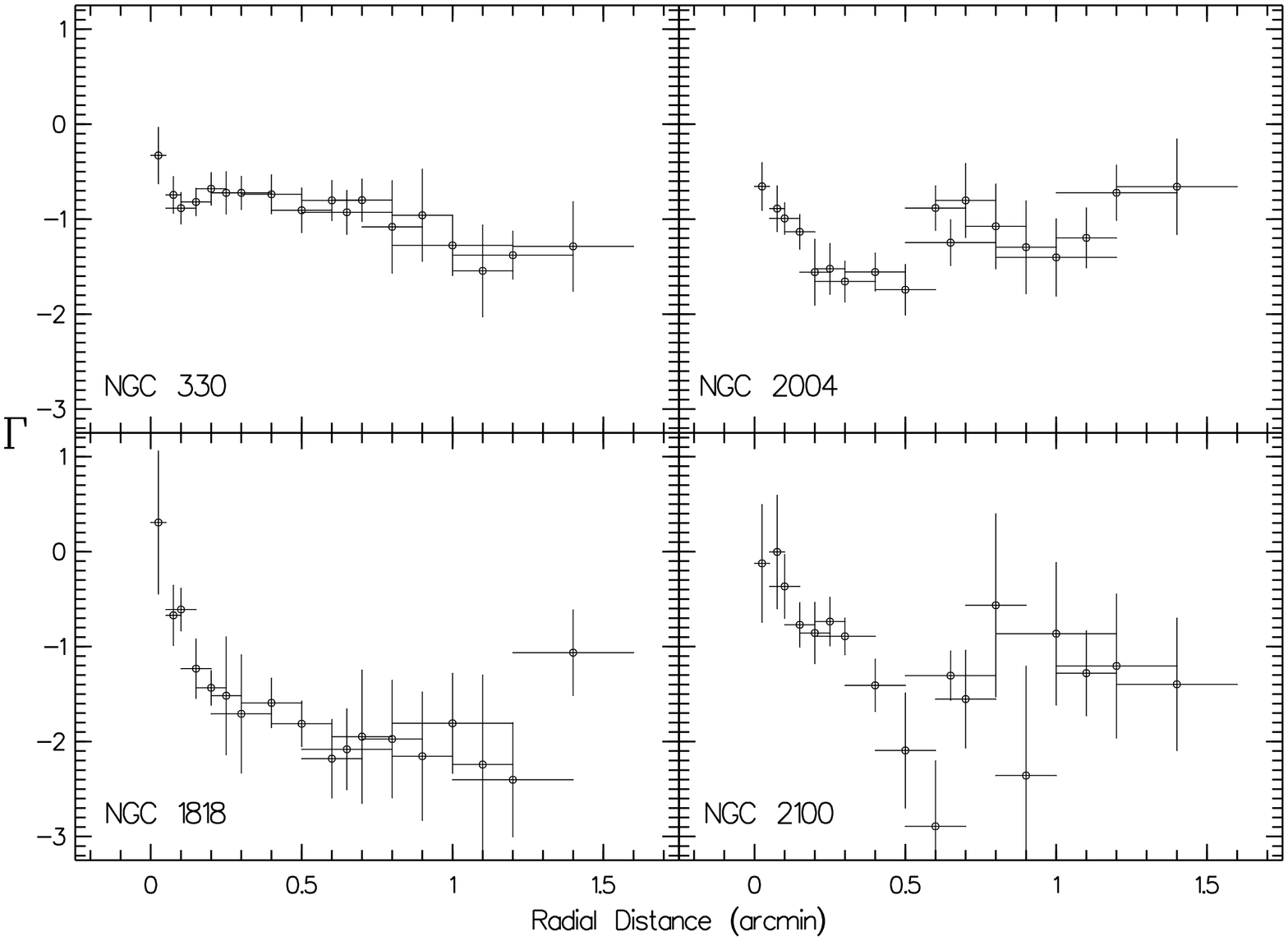,width=17.35truecm,angle=270}
}}
\caption{Radial dependence of the slopes of the differential MFs of the
clusters. The slopes have been estimated for MFs constructed with the
$second$ $method$.}
\label{mfslfig1}
\end{figure*}

As an example we present the radial MFs of the clusters in Fig.  
\ref{rdmffig} for four selected annuli (0\farcm0 - 0\farcm2, 0\farcm2 -
0\farcm4, 0\farcm4 - 0\farcm8 and larger than 0\farcm8). Each of these MFs
was corrected for incompleteness in the corresponding annulus. The changes
of the MF slopes of NGC 330 and NGC 2004 cannot be attributed to any mass
segregation, for NGC 330 due to the very small changes of $\Gamma$ and for
NGC 2004 because the slope does not show any trend. On the contrary, NGC
1818 and NGC 2100, as is shown from their MF slopes for the same annuli,
are found to be mass-segregated (also Fig. \ref{rdmffig}). We attempted to
check the MF slopes in a variety of rings around the centre of each
cluster in order to verify how much our results would be affected by a
different choice of the annuli. The results are shown in Fig.
\ref{mfslfig1}, where the radial MF slopes for various annuli are plotted
to the corresponding mean radial distance of each annulus. Specifically we
selected annuli spanning widths from 0\farcm05 for the inner ones to
0\farcm4 outwards. The horizontal ``error'' bars in Fig.  \ref{mfslfig1}
represent this width for every ring. These bars, as is shown, are
overlapping each other. This is because to estimate the slope of the
radial MF for each cluster we selected not only annuli with different
widths, but also at various distances from the centre of the cluster. The
selection of these annuli was random and subject to one only criterion:
Thinner annuli were selected toward the inner areas of the clusters and
wider outwards.

From the results in Fig. \ref{mfslfig1} one may conclude that all four
clusters in the sample seem to be mass-segregated, but in a different
degree and out to different radial distances (we call the radius within
which an important relation between MF slope and distance can be seen the
{\em ``radius of segregation''}). For NGC 1818 and NGC 2004 one may see
that there is a definite trend of the MF slope to steeper values up to
about 0\farcm6 and 0\farcm4 respectively.  The situation for NGC 330 is
more complicated since one may see a shallow trend of the MF slope up to
about 0\farcm5, which implies that there may be indeed a radial dependence
of the MF slope, which however seems to be very weak. On the other hand
this trend can be considered to exist up to the observed limits of the
cluster (see Sect. 4.2.3 below). For NGC 2100 the very noisy MF slopes
outwards cannot help us define a specific radius of mass segregation for
the whole of the cluster, except for the inner area. One, thus, could
argue that this cluster is mass-segregated throughout its observed extent.
We constructed the {\em cumulative} radial MFs of the cluster and we
verified this argument from the radial dependence of their slopes, where
also a small but steep radial dependence of the MF slope very close to its
centre (within 0\farcm2) was found (also faintly observed in Fig.
\ref{mfslfig1}).

In conclusion, as was found from the radial dependence of their MFs, NGC
1818, NGC 2004 and NGC 2100 are indeed mass-segregated. Specifically for
NGC 2100, there is a question whether the steep radial dependence of its
MF in the inner 0\farcm2 observed in Fig. \ref{mfslfig1} can be attributed
to a central mass segregation or not. The dependence of the MF of NGC 330
which can be seen in these figures for distances up to 0\farcm5 can be
questioned, due to the large errors of the MF slopes. Still, the
indications presented so far from the radial MFs of these clusters are
convincing that {\em all} the clusters in our sample are (to different
extents) mass-segregated, so from here on we treat all of them as such. We
applied the same tests to the slopes of the radial {\em cumulative} MFs
(where the statistics are much better) and we were able to verify the
observed mass segregation also from these slopes for all clusters.

\subsubsection{Comparison with previous results on the MF}

\noindent
{\em NGC 330}\\

\vspace*{-0.25truecm}
\noindent
Chiosi et al. (1995) investigated the IMF of NGC 330 as it is constructed
from its LF in connection to the selection of models concerning the mixing
procedures taking place in the interior of massive stars. For a mass range
$5 \leq m/{\rm M}{\solar}\leq 20$ the MF slope was found to be around the
Salpeter value for two of the three models (semiconvective and full
overshoot), while it was found to be flatter and equal to $\Gamma \simeq
-1$ assuming the diffuse overshoot model for the mixing in the interior of
massive stars. This value is very close to the MF slopes found by us for
this cluster in a narrower mass range (3 \lapprox\ $m/{\rm M}{\solar}$
\lapprox\ 9).

Concerning mass segregation in this cluster, Sirianni et al. (2002) who
present an extensive work on the low-mass end of the IMF of NGC 330 report
that the cluster is indeed segregated, since for stars more massive than
$\sim$ 1 M{\solar} the slope of the MFs measured at different radial
distances show a monotonic increase with increasing distance from the
centre of the cluster. We compared the relation of the MF slope to the
radial distance found by us with the corresponding relation by these
authors (their Fig. 8) and we found that indeed there is a similar trend
of the slope to steeper values outwards, which is, though, very shallow.
It should be noted that the MF slopes as were found by Sirianni et al.
(2002) are systematically steeper than ours by a small fraction, probably
due to the different lower mass range used by these authors (1 - 6
M{\solar}).

\vspace*{0.25truecm}
\noindent
{\em NGC 1818}\\

\vspace*{-0.25truecm} 
\noindent
This cluster may be one of the best observed in the LMC.  Hunter et al.  
(1997) report an overall MF slope of $\Gamma = -1.25 \pm 0.08$ for a mass
range between 0.85 and 9 M{\solar}, flatter than the MF of the cluster as
was found by us toward larger masses (4 - 11 M{\solar}). The same
authors found that the MF in the core of the cluster is slightly flatter
($\Gamma = -1.21 \pm 0.10$). The MF slopes found by Hunter et al. are
steeper than the one previously found by Will et al. (1995) for 2
\lapprox\ $m$/M{\solar} \lapprox\ 8, to be $\Gamma = -1.1 \pm 0.3$. More
recently de Grijs et al. (2002b)  presented the case of mass segregation
in NGC 1818 observed with HST and they report that Hunter et al. (1997),
as well as Santiago et al.  (2001) found systematically flatter MF slopes
than they did, probably due to the combination of different
mass-luminosity relations and a different treatment of the background
stellar population.

De Grijs et al. (2002b), in their investigation of the phenomenon of mass
segregation, applied a range of mass-luminosity relations (ML conversions)
to their observations for the construction of the MF of the cluster. We
compared our overall MF of NGC 1818 with the MFs of the cluster for the
three ML conversions used by these authors (their Fig. 4) and we found
that our MF for the mass range 3 \lapprox\ $m$/M{\solar} \lapprox\ 11
falls between de Grijs et al. MFs for two of the three empirical
conversions used: The one by Kroupa et al. (1993) and the one by Tout et
al. (1996), our MF being closer to the latter for both methods used for
its construction. In general our results are in good agreement with those
of de Grijs et al., with the MFs in both studies being steeper than those
previously reported. Our results are also in line with de Grijs et al.
concerning the phenomenon of mass segregation in NGC 1818 (see Sect. 5),
where they report that mass segregation becomes significant for masses
\gapprox\ 2.5 M{\solar}, out to at least 20\arcsec - 30\arcsec. They note
that the observed mass segregation is of primordial nature.

\vspace*{0.25truecm}
\noindent
{\em NGC 2004 \& NGC 2100}\\

\vspace*{-0.25truecm} 
\noindent 
These clusters have been the target of various investigations concerning
their young stellar content (for references see PhD Thesis by Grebel 1996
and Keller 2001). Still, no published results on their MFs are available
in the literature except for one case: A complete set of ground-based CCD
observations on five LMC clusters was presented by Sagar et al. (1991).
This set was used by Sagar \& Richtler (1991) for the determination of the
MF of these clusters, NGC 2004 and NGC 2100 being among them. It was found
that the MF slope of NGC 2004 for a mass range of 2 to 14 M{\solar} is
around $\Gamma \simeq -1.1$, between the two values of our observed MF
slopes (Fig. \ref{mfsfig}) for a similar mass range (3 - 14 M{\solar}).  
For the same mass range these authors found a flatter, but less reliable
MF slope for NGC 2100 ($\Gamma \simeq -0.8$). Indeed this slope is much
flatter than the one found by us (between $-1.1$ and $-1.2$).

Richtler et al. (1998) presented their preliminary results on the case
of mass segregation in NGC 2004 with the use of HST/WFPC2 observations. 
They checked the MF radial dependence at only two selected radii from
the centre of the cluster and they found that for the mass interval $1.3
< m/{\rm M}{\solar} < 4.27$ the MF slope indeed changes significantly.
Considering that we checked the MF radial dependence of the cluster 
for more annuli using different HST data, it is interesting to note that 
this result is in agreement with ours. Richtler et al. (1998) found 
that NGC 2004 is too young to be relaxed by two-body encounters, and thus 
they conclude that the observed mass segregation is primordial.

\section{Results on Mass Segregation} 

Since all clusters are found to be segregated (each to a different degree)
it would be useful to present some conclusive results on the phenomenon as
it was exhibited in every cluster, by combining the results of the
previous section (Sect 4.2) with those derived from the density profiles
and their dependence on the magnitude range (Sect. 3.2).

For NGC 330 it was found from star counts that within a radius of
0\farcm5, which might represent the radius of segregation (Sect. 4.2), a
significant fraction of stars have masses around 4 M{\solar}. In addition,
it is shown in Fig. \ref{denslp} that a significant change in the relation
of the slope of the density profile with the magnitude range appears
around $V \simeq$ 18 - 19 mag, which corresponds to 5.4 - 3.7 M{\solar}.
Still, the density profile dependence of this cluster on the corresponding
magnitude range was not easily detected and it seems to be rather weak.  
This also goes for the radial dependence of the MF slope of the cluster,
which gave us a rather uncertain radius of segregation. The best
segregated mass group for NGC 2004 seems to be also around 4 M{\solar},
within 0\farcm4 of the cluster, as was found from the radial dependence of
its MF slope.  This mass limit seems to be also in line with the results
of Fig. \ref{denslp}, where we see that mass segregation occurs at about V
$\simeq$ 18 - 19 mag (5.5 - 3.7 M{\solar}).  Still, this mass limit was
not so easily identifiable from the dependence of the density profiles on
the selected magnitude (mass) range, since this dependence is also faint
for NGC 2004.

Concerning NGC 1818 it seems that this cluster shows stronger
evidence of mass segregation down to 3 M{\solar} (V $\approx\ 20$ mag)  
and 6 M{\solar} (V $\approx\ 17.5$ mag), confined within about 0\farcm6
(as was found from the MF slope radial dependence of Fig.  
\ref{mfslfig1}). We also checked the differences between the regions
inside and outside the core radius of the cluster ($\sim$ 0\farcm2 - Table
1), where a central radial dependence of its MF slope can also be observed
(Fig. \ref{mfslfig1}) and we found that most of the stars with mass $\sim$
10 M{\solar} are confined within this radial distance. All these results 
seem to be in line with those on the magnitude dependence of the 
surface density profile slope, as is shown in Sect. 3.2, where we 
observed an almost continuous correlation between these parameters for NGC 
1818.

This dependence for NGC 2100 was found to be in general faint with a
significant change for stars more massive than about 9 - 7 M{\solar} ($V
\approx$ 17 - 18 mag) (see Sect. 3.2).  In addition, no specific radial
cut-off was found for the observed mass segregation, meaning that the
cluster is probably segregated throughout its extent (Sect.  4.2). Still,
a short but steep radial dependence of its MF was observed in the inner
0\farcm2 (Fig. \ref{mfslfig1}), and we found from star counts that the
most massive stars in the cluster ($\sim$ 13 M{\solar}) are located within
this radius.

\subsection{Binaries in Star Clusters}

Before exploring the primordial and/or dynamical mechanisms and their
possibilities of dominating the phenomenon of mass segregation we must not
exclude the contribution of binaries as a possible factor giving rise to a
false interpretation.

Binaries are known to exist in clusters and to dominate the first
dynamical period of their life. Still, binarity in studies of this kind is
usually not considered and if one includes the presence of binaries in the
MF construction then the MF slope becomes steeper (Kroupa 2000). Keller et
al. (2000) address the issue of the widening of the main sequence of the
clusters and they suggest the presence of binaries as a possible
explanation. They used near-infrared photometry of these clusters by
Keller (1999) in order to examine the ($V-K$) and ({\em F160BW} $-$ {\em
F555W}) colours and they revealed a number of systems that consist of a
binary pair of red supergiant and MS stars. They stressed the presence of
such systems in NGC 330 and NGC 1818.

Elson et al. (1998) investigated the binary populations in NGC 1818 and
they found that the full range of initial binary separations expected in
the cluster includes values up to 20 $\times$ 10$^{3}$ AU and that
resolved binaries with separations \gapprox\ 5 $\times$ 10$^{3}$ AU would
not survive in the core of the cluster to the present time. They also
found that there is no large population of resolved binaries in the outer
parts of the cluster. The limiting boundary between stars in soft and
those in hard binaries in the Elson et al. sample is 500 to 1000 AU.  The
same authors estimated from an examination of the lower MS a binary
fraction of around 20 - 30\% for the outer parts of the cluster and its
core respectively. If the binary fraction possesses a radial dependence
then this introduces a radial dependence on the luminosity function and it
would shift the LFs in Fig.  \ref{lfs} to lower $V$. However, we have
determined from simulations that in order to account in this way for the
displacement seen in Fig. \ref{lfs} for NGC 1818 a difference in binary
fraction between the inner and outer zones of the order of 50\% would be
required. Furthermore the sense of the difference would have to be such
that the binaries preferentially inhabit the cluster core. This should not
be the case, since it is theoretically shown that the binary frequency can
be reduced through stellar encounters in clusters and that the final
binary frequency depends on the stellar density in the cluster (Kroupa
1995). Dense environments, such as those in the clusters of our sample,
favour collisions and consequently there is a higher probability for
binary destruction to occur (Bonnell 1999). Recent models predict that
binary dissociation will take place in clusters denser than 10$^{3}$ stars
pc$^{-3}$ in time-scales shorter than 1 Myr destroying 50\% of
binaries wider than 100 AU (Bonnell 2000). Under these circumstances,
binarity should not modify any of our conclusions significantly.

\begin{table*}
\begin{center}
\caption{Estimated total number of stars $N$, total mass $M$ and the
corresponding relaxation time $t_{\rm rl}$ and stellar density $\rho$
within two characteristic radii of the clusters: the core radius ($r_{\rm
c}$) and the half-mass radius ($r_{\rm h}$), for the characteristic masses
of each cluster ($m_{\star}$) as found from the extrapolated MFs as
the median of the mass distribution$^{\ddagger}$. The errors reflect the
propagation of the Poisson uncertainties in the counts.}
\begin{tabular*}{\textwidth}[]{lc|rrrc|crrc}
\hline
&& \multicolumn{4}{c}{Assumed MF slope 
{\em unchanged} down to 0.1 M{\solar}}&
\multicolumn{4}{c}{Assumed MF slope
{\em unchanged} down to 2.0 M{\solar}}\\
&& \multicolumn{4}{c}{}&
\multicolumn{4}{c}{and {\em flat} for
2.0 {\gapprox} $m/{\rm M}{\solar}$ {\gapprox} 0.1}\\
\hline
Cluster&${r_{\rm c}}^{\dagger}$&\multicolumn{1}{c}{$N$}&
\multicolumn{1}{c}{$M$}&\multicolumn{1}{c}{$t_{\rm rlc}$}& 
$\log{\rho_{\rm c}}$&$N$&\multicolumn{1}{c}{$M$}&
\multicolumn{1}{c}{$t_{\rm rlc}$}& $\log{\rho_{\rm c}}$\\
Name&({\arcmin})&\multicolumn{1}{c}{($\times 10^{3}$)}&
\multicolumn{1}{c}{($10^{3}$ M{\solar})}&\multicolumn{1}{c}{(Myr)}&
(M{\solar} pc$^{-3}$)&\multicolumn{1}{c}{($\times 10^{3}$)}&
\multicolumn{1}{c}{($10^{3}$ M{\solar})}&
\multicolumn{1}{c}{(Myr)}&(M{\solar} pc$^{-3}$)\\
\hline
\hline
NGC 330 &0.15& 145.1 $\pm$ 0.4& 29.4 $\pm$ 0.1&   
80 $\pm$ 1& 2.71 $\pm$ 0.01& 0.41 $\pm$ 0.02& 1.6 $\pm$ 0.1& 
40 $\pm$ 1& 1.44 $\pm$ 0.05\\
NGC 2004&0.11&   5.9 $\pm$ 0.1&  2.5 $\pm$ 0.1&
22 $\pm$ 1& 2.04 $\pm$ 0.03& 0.15 $\pm$ 0.01& 0.7 $\pm$ 0.1& 
22 $\pm$ 1& 1.50 $\pm$ 0.09\\
NGC 1818&0.17& 196.2 $\pm$ 0.4& 40.6 $\pm$ 0.2&
94 $\pm$ 1& 2.68 $\pm$ 0.00& 0.58 $\pm$ 0.02& 2.0 $\pm$ 0.1& 
43 $\pm$ 1& 1.38 $\pm$ 0.05\\
NGC 2100&0.08&   2.7 $\pm$ 0.1&  1.4 $\pm$ 0.1&
12 $\pm$ 1& 2.19 $\pm$ 0.05& 0.10 $\pm$ 0.01& 0.6 $\pm$ 0.1& 
15 $\pm$ 1& 1.82 $\pm$ 0.11\\
\hline
&${r_{\rm h}}$&\multicolumn{1}{c}{$N$}&
\multicolumn{1}{c}{$M$}&\multicolumn{1}{c}{$t_{\rm rlh}$}& 
$\log{\rho_{\rm h}}$&$N$&\multicolumn{1}{c}{$M$}&
\multicolumn{1}{c}{$t_{\rm rlh}$}& $\log{\rho_{\rm h}}$\\
&({\arcmin})&\multicolumn{1}{c}{($\times 10^{3}$)}&
\multicolumn{1}{c}{($10^{3}$ M{\solar})}&\multicolumn{1}{c}{(Myr)}&
(M{\solar} pc$^{-3}$)&\multicolumn{1}{c}{($\times 10^{3}$)}&
\multicolumn{1}{c}{($10^{3}$ M{\solar})}&
\multicolumn{1}{c}{(Myr)}&(M{\solar} pc$^{-3}$)\\
\hline
NGC 330 &0.26&1743.2 $\pm$ 1.3 &290.7 $\pm$ 0.3&   
356 $\pm$ 1&2.98 $\pm$ 0.00& 0.84 $\pm$ 0.03 &3.2 $\pm$ 0.1 & 
87 $\pm$ 2&1.03  $\pm$ 0.04\\
NGC 2004&0.22& 11.0 $\pm$ 0.1& 5.4 $\pm$ 0.1&
52 $\pm$ 1& 1.47 $\pm$ 0.02& 0.48 $\pm$ 0.02& 2.1 $\pm$ 0.1& 
52 $\pm$ 2& 1.06 $\pm$ 0.05\\
NGC 1818&0.27& 97.4 $\pm$ 0.3& 24.9 $\pm$ 0.1&
124 $\pm$ 1& 1.87 $\pm$ 0.01& 0.78 $\pm$ 0.03& 2.8 $\pm$ 0.1& 
77 $\pm$ 2& 0.92 $\pm$ 0.04\\
NGC 2100&0.25& 9.5 $\pm$ 0.1& 6.0 $\pm$ 0.1&
64 $\pm$ 1& 1.35 $\pm$ 0.02& 0.61 $\pm$ 0.02& 2.8 $\pm$ 0.1& 
65 $\pm$ 2&  1.02 $\pm$ 0.05\\
\hline
\end{tabular*}
\parbox[l]{\textwidth}{
$^{\ddagger}$ The values of $m_{\star}$ that were found are 
(in M{\solar}) 0.98, 1.23, 1.09 and 1.23 for NGC 330, NGC 2004,
NGC 1818 and NGC 2100 respectively.\\ 
$^{\dagger}$ From Mackey \& Gilmore (2003a,b)}
\end{center}
\label{mypars}
\end{table*}

\section{Discussion}

\subsection{Relaxation Times of the Clusters}

As we discussed in Sect. 1, the observed mass segregation could have
either a dynamical or primordial basis. Dynamical evolution is known to be
achieved through two-body encounters and consequently its extent is
related to the age of the stellar population. The characteristic
time-scales involved in the relaxation of the system are: Its crossing
time ($t_{\rm cr}$), which is the time needed by a star to move across the
system, and its two-body relaxation time ($t_{\rm rl}$), which is the time
needed by the stellar encounters to redistribute their energies, setting
to a near-Maxwellian velocity distribution. The latter timescale is
extremely significant in the case of mass segregation, because one may
test if it is of dynamical origin (when the evolution time of the system,
$\tau$, is longer than $t_{\rm rl}$) or not.  The relaxation time of a
cluster can be characterised by its {\em ``Median Relaxation Time''}, i.e.
the time after which gravitational encounters of stars have caused the
system to equilibrate to a state independent of the original stellar
orbits; it can be expressed as (Binney \& Tremaine 1987):
\beq
t_{\rm rl} = \frac{6.63 \times 10^{8}}{\ln{0.4N}}
{\left ( \frac{M}{10^{5}{\rm M}{\solar}} \right )}^{1/2}
\left ( \frac{{\rm M}{\solar}}{m_{\star}} \right )
{\left ( \frac{\cal{R}}{\rm pc} \right )}^{3/2} {\rm yr}
\label{trel}
\eeq
where $M$ is the total mass of the system within some characteristic
radius $\cal{R}$, $m_{\star}$ is a characteristic stellar mass (the median
of the observed mass distribution) and $N$ is the corresponding total
number of stars of the system. A characteristic radius of a stellar system
is its half-mass radius ($r_{\rm h}$) and one can use this radius in Eq.
(\ref{trel}) to estimate the {\em ``Half-mass Relaxation Time''} of the
system (e.g. Portegies Zwart et al. 2002). We also estimated the
relaxation time of the systems within their core radii ($r_{\rm c}$). In
order to derive the median relaxation time we need to have an estimation
of the characteristic radii above, as well as the total mass of the system
within these radii and the corresponding characteristic masses $m_{\star}$
and total number of stars $N$.

Mackey \& Gilmore (2003a,b) recently published surface brightness profiles
for a sample of 53 LMC and 10 SMC rich star clusters, and they derived
structural parameters for each cluster using their detailed profiles made
with the use of two-colour observations from HST. Their results thus
consist of a coherent sample of parametric values for the dynamical
properties of clusters in the MCs (see Table 1). These values were
estimated by the fit of the model introduced by Elson et al. (1987) to
their surface brightness profiles. All the clusters of our sample are
included in the catalog of Mackey \& Gilmore, so we can use the core radii
for our clusters as they were estimated by these authors.

We estimated the radius where half of the observed total mass (corrected
for incompleteness and field contamination) of the clusters is confined.
We refer to this radius as the {\em observed} half-mass radius. If we
assume that the low-mass stars are uniformly distributed over the area
covered by each cluster, then the radius where half of the total mass is
included should not be far from this estimated value. The values of
$r_{\rm c}$ and the observed $r_{\rm h}$ are given in the second column of
Table 3.

The total number of stars $N$, the corresponding total mass $M$ and the
characteristic mass $m_{\star}$ within the core and half-mass radius of
each cluster can be evaluated by their MFs. The main uncertainty is the
low end of the mass function which is not known because of the detection
limit.  In principle there is no reason to believe that the MFs of the
clusters in our sample have different low-mass cut-offs. In addition this
lower mass limit, $m_{\rm low}$ is very difficult to constrain, especially
for distant and dense clusters like those of our sample, so we can only
make some reasonable guess. Brandl et al. (1999) using observations from
ANTU/ISAAC have revealed that in the starburst region NGC 3603 of our
galaxy there are many of sub-solar mass stars down to 0.1 M{\solar}.
According to these authors this result agrees with studies in the solar
neighbourhood, which also indicate that $m_{\rm low} \simeq$ 0.1
M{\solar}. Based on these results we extrapolated the observed MFs of the
clusters down to this mass limit.

Since the clusters were found to be mass segregated, meaning that their
MFs have different slopes within different radial distances, we did not
extrapolate the overall MFs but separately the MF within $r_{\rm c}$ and
the one within the observed $r_{\rm h}$, using in every case the
corresponding slope, which is assumed to be the one derived from the most
complete observed low-mass bins, because, as we already noted in Sect.
4.2, the MFs may not be represented in total by a single power-law due to
the change of their slopes toward different end masses (see Figs.
\ref{mfsfig} and \ref{rdmffig}). The next question is if the MF follows
the same slope down to the limit of 0.1 M{\solar}. First, the
extrapolations mentioned above were done assuming that indeed the low-mass
MF slope down to 0.1 M{\solar} is the same as the MF slope of the more
massive stars. Sirianni et al. (2000) constructed the IMF of R 136 with
the use of deep broadband $V$ and $I$ HST/WFPC2 observations and they
report that, after correcting for incompleteness, the IMF shows a {\em
definite} flattening below $\simeq$ 2 M{\solar}. Thus, we also took these
results into consideration.

Therefore, in order to estimate the total mass of the clusters we
extrapolated their observed MFs, as they were constructed within both
their core and half-mass radii, under two assumptions: i) That the MF
slope remains unchanged down to 0.1 ${\rm M}_{\solar}$ and ii) that this
slope remains unchanged down to 2.0 ${\rm M}_{\solar}$, but becomes flat
for masses down to 0.1 ${\rm M}_{\solar}$. The results of the estimations
above are given in Table 3, where the total number of stars and the
corresponding masses for both assumptions concerning the MF slope are
given (Cols. 3,4 and 7,8 respectively). The derived relaxation times are
given in Col. 5 for the first assumption and in Col. 9 for the second. We
were also able to estimate the corresponding stellar densities within both
radii for both assumptions on the MF extrapolation, which are given in
Cols. 6 and 11 respectively.

\subsection{Dynamical or Primordial Mass Segregation?}

The results of Table 3 mostly demonstrate the sensitivity of the
calculations to the assumed MF slope toward the low-mass end. Still,
considering the two cases assumed for the MF extrapolation to lower masses
one can safely reach some interesting conclusions, such as that the
clusters do not appear to be dynamically relaxed within their observed
half-mass radius, since their evolution times are shorter
than their relaxation time-scales. Concerning their relaxation within
their core radii the results are rather different. If we assume that
the MF follows the same slope down to 0.1 M{\solar} then NGC 330 and NGC
1818 do not seem to have had the time to reach relaxation in their cores.
This is not the case for NGC 2004 and NGC 2100, which appear 
to be almost (if not completely) relaxed in their cores with both
assumptions for their MFs, which makes this result more solid. It is worth
noting that if the MFs of NGC 330 and NGC 1818 are assumed to be flat for 
stars $<$ 2 M{\solar}, then the cores of these clusters appear to be 
relaxed now.

To make a simple check on the accuracy of our MF extrapolations we
performed the same calculations for the whole area of each cluster and we
found the total mass of each cluster from its extrapolated MF. The
parametric dynamical masses of the clusters as they were found by Mackey
\& Gilmore (Col. 8 of Table 1) can be used as a guideline, since we can
compare them with our estimated masses of the clusters. We found that the
parametric mass values favour the extrapolation of the MF that assumes the
same slope down to 0.1 M{\solar} for most of the clusters (no flattening
below 2 M{\solar}). Specifically the masses of NGC 2100 and NGC 2004 as
estimated from their extrapolated MFs are in excellent agreement with
their dynamical masses if they include stars down to 0.1 M{\solar}. Our
estimation of the mass of NGC 330 was found to be within the limits of its
parametric value, but for NGC 1818 we found that in order to have
comparable values for the parametric and observed mass of the cluster its
MF should be extrapolated (with the same slope) only down to about 1
M{\solar}. Still, for the sake of consistency for the estimations that
follow, we assumed extrapolation of the MF down to 0.1 M{\solar} for all
clusters.

It would be interesting to know the relaxation time also within the radius
of segregation for each cluster. Therefore, we performed similar
calculations to the ones for the relaxation times within $r_{\rm c}$ and
$r_{\rm h}$ and we defined the time needed for the clusters to relax
within their radii of segregation, meaning the radial distances where we
detected a significant difference in their MF slopes (Sect.  4.2.2). For
NGC 330 we found that the cluster needs 2 to 7 $\times\ 10^{8}$ yr to
relax within 0\farcm5, i.e. the distance where a hint of segregation was
found. For NGC 1818 a relaxation time within its segregation radius of
0\farcm6 was found between 2 and 6 $\times\ 10^{8}$ yr. The relaxation
time of NGC 2004 within its segregation radius (0\farcm4) was found to be
around $10^{8}$ yr (with both assumptions on its MF extrapolation). The
radius of 0\farcm2 in NGC 2100, where a central mass segregation was
observed is fairly close to its observed half-mass radius, so the results
are similar with a relaxation time between 50 and 53 Myr. If we assume
that this cluster is segregated througout its extent (see Sect. 4.2.2)
then it would need 6 - 7 $\times\ 10^{8}$ yr to relax dynamically, which
is much longer than its age.

The range of these values for the relaxation time of the clusters is
mostly due to the assumption adopted for the low-mass end of the MFs. The
derived number of stars and the corresponding masses in turn were mostly
affected by the assumed MF slope toward this end.  So one should keep in
mind that all these results can give us only a rough idea on the dynamical
status of the clusters under consideration. Stolte et al. (2002),
investigating the mass segregation of the Arches cluster, make a very
interesting comment concerning such approaches. They note that the
observed relaxation time does not properly trace the cluster's initial
conditions, but reflects its current dynamical state and they conclude
that distinction between primordial and dynamical mass segregation is not
possible for the cluster. In any case, one can apply simple tests, which
are based on theory, so that conclusions can be derived on the
equipartition among stars of different masses as they are observed in a
star cluster. We describe these tests as follows.

Concerning the equipartition time-scale of clusters, simulations by
Inagaki \& Saslaw (1985) predict that the heaviest stars are located near
the center fairly soon. This is in line with Chernoff \& Weinberg (1990),
who show that during multi-mass collapse equipartition for heavy-mass
groups occurs {\em before} (dynamical) mass segregation takes place in the
cluster as a whole.  Spitzer (1969) showed that in a two-component stellar
system, if the total mass of the heavier stars exceeds a critical value,
equipartition with the lighter stars becomes impossible. If ${\rm m}_{1}$
and ${\rm m}_{2}$ are the two mass groups in the cluster with ${\rm
m}_{2}/{\rm m}_{1}\gg 1$ and M$_{i}$ is the total mass of stars in the
group of mass ${\rm m}_{i}$ then equipartition between the two mass groups
can be achieved only if:
\beq 
\frac{{\rm M}_{2}}{{\rm M}_{1}} < 0.16 {\left( \frac{{\rm m}_{1}}{{\rm 
m}_{2}}\right )}^{3/2} 
\label{sptzcnt}
\eeq

For multicomponent systems Inagaki \& Saslaw (1985) showed that conditions
for global equipartition do not generally arise in stellar systems with a
range of masses. These conditions are so restrictive that they are
unlikely to apply to realistic clusters. Furthermore, Vishniac (1978)
found that the generalisation of Eq. (\ref{sptzcnt}) for a system with a
specific mass function is the requirement for a MF index $\alpha$
\gapprox\ 3.5 (Salpeter's index: $\alpha = 2.35$) and Inagaki \& Saslaw
(1985) found that {\em local equipartition may occur temporarily near the
centre among all mass species only if the mass spectrum is even steeper}.  
If the mass spectrum is not very steep only the massive stars can be in
approximate equipartition near the centre of the cluster. These authors
considering a system of three mass groups ${\rm m}_{3}>{\rm m}_{2}>{\rm
m}_{1}$, expanded the conditions of Eq. (\ref{sptzcnt}) to the following:
\beq 
\frac{{\rm M}_{3}}{{\rm M}_{1}} < 0.38 {\left( \frac{{\rm m}_{3}}{{\rm 
m}_{2}}\right )}^{-3/2}~~~{\rm and}~~~ \frac{{\rm M}_{2}}{{\rm M}_{1}} < 
0.38 {\left( \frac{{\rm m}_{2}}{{\rm m}_{1}}\right )}^{-3/2} 
\label{ingkcnt}
\eeq
the first being the condition for equipartition between the second and the
third mass group and the second being the condition for equipartition
between the first and the second group (comparable to Eq.
(\ref{sptzcnt})). They conclude that it is easier to attain equipartition
between the massive stars than between the lighter ones.

From the information provided in the previous paragraphs one may
conclude that: (i) The clusters of our sample are not expected to
demonstrate any equipartition throughout their area and between all their
mass groups, due to their rather flat MFs. (ii) The only stars in 
equipartition (if any) should be the most massive ones toward the 
centre of the clusters. Considering these two
points it seems rather interesting to test the conditions of Eq.
(\ref{ingkcnt}) on our clusters using three of their observed massive
stellar groups. For this test we used three indicative mass groups from 
the most massive bins of the MFs of the clusters and we estimated the
corresponding total masses for every group using the stellar numbers as
they were found from these MFs. We applied the calculations on the MFs
of the clusters within their core radii, observed half-mass radii, as
well as within their segregation radii. In all cases we used the stars
observed to be more massive than 5 M{\solar}. We tested the conditions
of Eq. (\ref{ingkcnt}) for all the possible combinations of massive MF
bins (stellar groups) that satisfy the inequality ${\rm m}_{3}>{\rm
m}_{2}>{\rm m}_{1}$.

For NGC 330 we used combinations between 6 mass groups with masses up to
$\sim$ 8.5 M{\solar}. NGC 2004 and NGC 2100 have a wider mass range and so
the conditions for equipartition were tested among 9 mass groups more
massive than $\sim$ 5 M{\solar}, with upper limit $\sim$ 13 M{\solar}. NGC
1818 was tested for 7 mass groups up to $\sim$ 10 M{\solar}. In all cases
it was found that the conditions of Eq. (\ref{ingkcnt}) could not be
satisfied at any radius for any mass group for all clusters. It is worth
noting that the used stellar mass groups and their corresponding total
masses are purely observed quantities, so these results are free from any
assumption or model. In conclusion, it seems that in the clusters of our
sample no equipartition took place between the massive stars within their
cores, half-mass radii, or their radii of segregation.

The simulations by Inagaki \& Saslaw (1985) suggest that the tendency of
massive stars to reach equipartition toward the centre is expected to
occur early on and increasing mass segregation later deletes any signature
of this equipartition. Could this be the case for our clusters? If so we
would not be able to verify any equipartition between the massive groups
of stars because mass segregation has erased any trace of it.  In a more
recent work Bonnell \& Davies (1998) determine that the massive stars do
segregate very fast but only if they are born within the central 10\% to
20\% of all stars.  Bonnell (1999) concludes that dynamical evolution and
mass segregation can be achieved fast under the condition that massive
stars are formed within a specific volume of the cluster. Still,
equipartition and mass stratification is expected to be accomplished in
roughly one relaxation time (e.g. Spitzer \& Shull 1975), and the
estimated relaxation times of Table 3 do not favour the possibility of
completed mass segregation, which would have destroyed any initial
equipartition among the massive stars, unless we have seriously
underestimated the slopes of the MFs of our clusters. This clearly implies
that {\em the observed mass segregation seems to be a primordial effect,
related to the loci where massive stars have been formed}.

Recent results on the Orion Nebula Cluster (ONC) and the Trapezium cluster
in its core (McCaughrean \& Stauffer 1994; Hillenbrand 1997; Hillenbrand
\& Hartmann 1998) have given us the opportunity to reconsider the initial
dynamical conditions of star formation. Some of these results have been
taken as an indication that star formation is bimodal, with high-mass
stars forming in a different way from low-mass stars. Bonnell \& Davies
(1998) constrained their models in order to answer the question if the
bias of the young stellar population in the ONC towards high-mass stars in
the central region of the cluster is due to the formation of massive stars
in their present locations or to subsequent dynamical evolution. They
found that for the present conditions of the cluster to be reproduced it 
must have been somewhat mass-segregated when it formed.

The star-formation scenario proposed by Bonnell \& Davies (1998) seems to
agree with the one by Murray \& Lin (1996), since the centrally located
protostellar clumps grow either through accretion from the residual gas or
through coagulation with other fragments. In the later case the
collisional timescale for the Trapezium cluster is of the order of 3
crossing times of the central region or one third of the crossing time of
the whole cluster, which is not supported by the observations. This leaves
only the alternative of subsequent accretion of the residual gas present
in the cluster to be more likely to have produced the observed range of
stellar masses.  Models of accretion in small stellar clusters (Bonnell et
al. 1997) showed that the stars near the centre of the cluster have the
highest accretion rates. Thus the subsequent accretion of material
produces the most massive stars in the centre of the cluster and initial
mass segregation takes place. It is worth noting that both scenarios cited
here seem to favour environments with high central density for primordial
mass segregation to occur.

\section{Conclusions}

The radial distribution of the cluster density in various magnitude ranges
and its MF within different radial distances are two useful diagnostic
tools for mass segregation. Both require 1) very good resolution towards
the cluster centre in order to have accurate radial density distributions
for various magnitude ranges and 2) very good photometry for accurate MFs
in rings for various radial distances in order to find at which distance
mass segregation leads to a change in the slope of the MF. Our HST/WFPC2
observations on young clusters in the MCs provide a suitable data
set for the application of these tools. The conclusions of this
investigation on the phenomenon of mass segregation in these clusters are
the following.

We searched for mass segregation in a sample of four young star clusters
in the MCs, NGC 1818, NGC 2004 and NGC 2100 in the LMC and NGC 330 in the
SMC. These clusters are among the younger MC clusters and they vary in
age, metallicity and other parameters. Mass segregation was exhibited in
all four of them, but in a variety of related parameters, as well as of
degrees of significance. From the dependence of their density profiles on
the selected magnitude range it was found that NGC 1818 exhibits strongly
the phenomenon of mass segregation with stars segregated almost in the
whole of the observed magnitude range, while NGC 330 and NGC 2100 show a
rather weak dependence and it is important only for stars around $V\simeq
18$ mag ($\sim$ 4 M{\solar}).  For NGC 2004 there is an even weaker
dependence also for $V \sim 18$ mag.

We constructed the mass functions (MFs) of the clusters for the observed
range of magnitudes in order to check for any radial dependence of their
slopes as the signature of mass segregation. This construction was
performed with two methods. First by directly counting stars between
evolutionary tracks according to their positions in the HRD and second by
translating their luminosities into masses using mass-luminosity relations
and then constructing the distribution of the estimated masses. The
$first$ $method$ which is more accurate concerning the shape of the MF but
less statistically significant due to the lower numbers, was used to test
the results of the $second$ $method$, which is more statistically complete
but more model dependent.

We constructed the MFs of the clusters considering the dependence of their
shape to the chosen binning for counting stars, the correction for
incompleteness and the field contamination. The use of a mass-luminosity
relation can affect the shape of the resulting MF (which tends to be
flatter for older population models). Still, it was found that the models
used for the $second$ $method$ were sufficient, since the overall MFs do
not differ significantly from one method to the other and they have slopes
clustered around a Salpeter IMF for a mass range of $\sim$ 3 - 9 M{\solar}
for NGC 330, 3 - 14 M{\solar} for NGC 2004, 4 - 11 M{\solar} for NGC 1818
and 4 - 14 M{\solar} for NGC 2100. It should be noted that not all of them
may be represented by a single-power law for the whole mass range
mentioned.

The construction of the radial MFs for the investigation of their slopes
proved to be a rather complicated procedure due to several factors.
Counting stars for the construction of the MF of a cluster in a specific
ring around the centre can produce statistical problems due to the low
numbers. Thus, the $first$ $method$ was excluded from the investigation of
the radial dependence of the MF slope of the clusters, since by
definition, it can rely on a smaller sample of stars.  NGC 1818 and NGC
2004 as was found from the radial dependence of their MFs are 
mass-segregated at radial distances of about 0\farcm6 and 0\farcm4
respectively. NGC 2100 was also found to be mass-segregated probably
throughout its observed extent, but a steep radial dependence of its
MF in the inner 0\farcm2 was also observed and this can be assigned to a
local central mass segregation of the most massive group of stars ($\sim$
13 M{\solar}) as was found from the relative numbers of stars inside and
outside this radial limit. This was also the case for NGC 1818 where 99\%
of the stars with masses around 10 M{\solar} were found concentrated
within about 0\farcm2 from its centre. NGC 330 was also found to be
segregated from the radial dependence of its MF, but in a less pronounced
way up to a distance of $\sim$ 0\farcm5, with the best segregated mass
group to be around 4 M{\solar}. This cluster could also be considered as
mass segregated throughout its extent.

We estimated the median relaxation time of the clusters in their core and
observed half-mass radii, as well as in their ``radii of segregation'' as
derived from the radial dependence of their MF slopes. This estimation was
based on assumptions on the slope, and on the low-mass limit, and we found
that the clusters in this sample are not dynamically relaxed within these
radii, except for NGC 2004 and NGC 2100, which were found to be very close
to relaxation only in their cores. Furthermore, the test of conditions for
equipartition among three massive groups (Inagaki \& Saslaw 1985) in our
clusters showed that {\em no equipartition has taken place} in any cluster
at any of the above radii for stars with masses larger than 5 M{\solar}.
This clearly suggests that the observed mass segregation is of primordial
nature. We present a discussion where we highlight the most important
aspects of theories of star formation predicting the phenomenon of initial
mass segregation (Sect. 6.2).

In general it seems that the diagnostic tools applied here, which are
those most commonly used for the investigation of mass segregation, are
rather sensitive to selection effects, such as the selection of the radial
distances for which the MF slopes are computed, the binning selected for
the construction of the MF, the selection of the mass range where the
slopes are estimated, etc. While the presence of mass segregation in our
sample of clusters is certainly a robust result, the description of the
phenomenon can be only qualitative.  Specifically, we observe that the
clusters of our sample exhibit mass segregation in different ways, but to
describe these differences we would need to define a new parameter. This
parameter may be called ``degree of mass segregation'', and it should be a
function of parameters involved in the magnitude dependence of the density
profile and the radial dependence of the MF slope for each cluster. It is
not yet possible to define such a parameter with the use of the available
tools, due to the limitations mentioned above.  Gouliermis \& de Boer
(2003) recently proposed a new method for the detection of stellar
stratification in star clusters. They present a self-consistent,
model-free tool for the investigation of the phenomenon, which can also be
used to quantify the degree of mass segregation in star clusters in a
uniform manner. This tool would probably provide a coherent scheme for the
description of stellar stratification in star clusters, so that global and
comparable results can be achieved.

\begin{acknowledgements}

S. C. Keller acknowledges the support of an Australian Postgraduate
Research scholarship and a grant from the DIST {\it{Hubble Space
Telescope}} research fund. This work is based on observations with the
NASA/ESA {\em Hubble Space Telescope}, obtained at the Space Telescope
Science Institute, which is operated by the Association of Universities
for Research in Astronomy, Inc., (AURA), under NASA Contract NAS 5-26555.

\end{acknowledgements}


\end{document}